\pdfoutput=1
\documentclass[journal]{IEEEtran}
\usepackage{amsmath}
\usepackage{amsthm}
\usepackage{amssymb}
\usepackage{subfigure}
\newtheorem{question}{\bf Question}
\newtheorem{example}{\bf Example}
\newtheorem{definition}{\bf Definition}
\newtheorem{game}{\bf Game}

\newtheorem{observation}{\bf Observation}
\newtheorem{proposition}{\bf Proposition}
\newtheorem{thm}{\bf Theorem}

\usepackage{tcolorbox}
\newtheorem{problem}{\bf Problem}
\usepackage{cite}
\usepackage{bm}
\usepackage{amsfonts}
\usepackage{cases}
\usepackage[linesnumbered,ruled,vlined]{algorithm2e}
\usepackage{svg}
\usepackage{graphicx}
\usepackage{modroman}
\DeclareMathOperator*{\argmax}{argmax}

\ifCLASSINFOpdf
\else
\fi
\begin{document}
%
\title{Machine Learning Model Trading with Verification under Information Asymmetry}
%
%
%

\author{Xiang~Li,
        Jianwei~Huang$^*$,~\IEEEmembership{Fellow,~IEEE,}
        Kai~Yang$^*$,~\IEEEmembership{Senior Member,~IEEE,}
        and~Chenyou~Fan.
\thanks{$^*$Jianwei Huang and Kai Yang are co-corresponding authors of this paper.}
\thanks{Xiang Li is with Shenzhen Institute of Artificial Intelligence and Robotics for Society (AIRS) and School of Science and Engineering (SSE), The Chinese University of Hong Kong, Shenzhen (Email: xiangli2@link.cuhk.edu.cn). Jianwei Huang is with SSE, AIRS, Shenzhen Key Laboratory of Crowd Intelligence Empowered Low-Carbon Energy Network, and CSIJRI Joint Research Centre on Smart Energy Storage, The Chinese University of Hong Kong, Shenzhen, Guangdong, 518172, P.R. China (Email: jianweihuang@cuhk.edu.cn). Kai Yang is with Department of Computer Science and Technology, Tongji University, Key Laboratory of Embedded System and Service Computing Ministry of Education, and Shanghai Research Institute for Intelligent Autonomous Systems (Email: kaiyang@tongji.edu.cn). Chenyou Fan is with South China Normal University (Email: fanchenyou@gmail.com).}
\thanks{Part of analysis and results appeared in IEEE ICC 2023 Conference \cite{Li2023}.}
}

%
%

\markboth{IEEE/ACM Transactions on Networking}%
{Shell \MakeLowercase{\textit{et al.}}: Bare Demo of IEEEtran.cls for IEEE Journals}
%



\maketitle

\begin{abstract}
Machine learning (ML) model trading, known for its role in protecting data privacy, faces a major challenge: information asymmetry. This issue can lead to model deception, a problem that current literature has not fully solved,  where the seller misrepresents model performance to earn more. We propose a game-theoretic approach, adding a verification step in the ML model market that lets buyers check model quality before buying. However, this method can be expensive and offers imperfect information, making it harder for buyers to decide. Our analysis reveals that a seller might probabilistically conduct model deception considering the chance of model verification. This deception probability decreases with the verification accuracy and increases with the verification cost. To maximize seller payoff, we further design optimal pricing schemes accounting for heterogeneous buyers' strategic behaviors. Interestingly, we find that reducing information asymmetry benefits both the seller and buyer. Meanwhile, protecting buyer order information doesn’t improve the payoff for the buyer or the seller. These findings highlight the importance of reducing information asymmetry in ML model trading and open new directions for future research.
\end{abstract}


%
\IEEEpeerreviewmaketitle
\section{Introduction}
\IEEEPARstart{M}{achine} learning (ML) has become increasingly crucial in many data-driven applications. Accordingly, ML model trading is also growing in popularity, as evidenced by various model markets, e.g.,  Modzy \cite{Modzy} and StreamML \cite{StreamML}. In particular, unlike data trading, model trading can effectively mitigate privacy concerns by avoiding direct access to data. Moreover, the model buyer can readily deploy purchased ML models, eliminating the need for additional data processing.

While recent studies (e.g., \cite{chen2019towards,agarwal2019marketplace,liu2020dealera,sunpeng}) have made initial attempts to analyze ML model markets, they typically assume a complete information scenario, i.e., model quality is transparent to both the buyer and seller. In practice, however, the buyer often lacks knowledge of the actual model quality before purchase and deployment, leading to \emph{information asymmetry}.\begin{figure}[ht]
    \centering
    \includegraphics[width=0.42\textwidth]{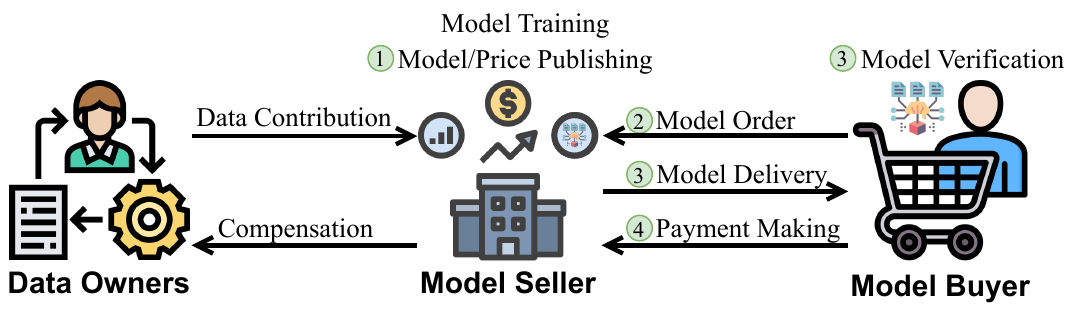}\vspace{-8pt}
    \caption{The Framework of ML Model Trading with Verification (MTV)}
    \label{framework}\vspace{-15pt}
\end{figure} The self-interested seller may misrepresent model quality to maximize payoff, even given future consequences like reputational costs \cite{9973974}, as long as the benefits outweigh these costs. This can result in the buyer overpaying for low-quality models and facing detrimental effects. For example, in bridge safety monitoring, a low-quality detection model producing false predictions could have disastrous outcomes. To avoid potential losses, buyers may thus opt to purchase only the lowest-quality models \cite{akerlof1978market} or refrain from participating in trading altogether, hindering the development of model markets.


To deal with the challenge of information asymmetry, we propose the ML \underline{M}odel \underline{T}rading with \underline{V}erification (MTV) framework, as illustrated in Figure \ref{framework}. The MTV framework enables the buyer to verify model quality with their test data (e.g., through a test interface as in Aimodelplace \cite{Modelplace}) before purchasing, reducing the information asymmetry regarding model quality. However, the verification cost associated with data collection and the imperfect verification arising from sampling bias of test data \cite{kramer2013importance} may discourage the buyer from verification. This motivates us to study the first key question:

\begin{question}
\label{q1}
How should buyers conduct model verification?
\end{question}

The unique characteristics of ML models, along with intertwined factors in model trading, present challenges for solving Question \ref{q1}. Specifically, due to the probabilistic nature of model predictions \cite{friedman2001elements}, assessing model quality with verification results and determining corresponding acceptance criteria are highly challenging for the buyer. Moreover, the cost-effectiveness of verification depends not only on the attributes of both the buyer and seller (e.g., test data size and model quality versions) but also on their interactions, which further complicates the buyer's strategic verification decision.

Given the above issues, we theoretically quantify the results of model verification to tackle Question \ref{q1}. In particular, we evaluate the buyer's verification accuracy in a probabilistic manner, taking into account the test data size and purchased model performance. By analyzing the buyer’s expected payoff under different conditions, we derive optimal decisions regarding model verification. Building upon these, we proceed to investigate our second key question in this paper:

\begin{question}
\label{q2}
How does the verification mechanism affect the payoffs of both the buyer and seller in ML model trading?
\end{question}

The imperfect information of trading strategies between the buyer and seller introduces difficulty in assessing individual payoffs. Meanwhile, as the number of model versions (with different quality levels) grows, deriving the optimal decision and equilibrium payoff becomes progressively complicated. Given that improvements in ML model generalization error (quality) exhibit a steep power-law scaling with training data size (cost), e.g., \cite{hestness2017deep,amari1993universal,amari1992four}, we simplify the equilibrium analysis in model trading. Specifically, by exploiting the power-law feature of learning curves, we reduce the seller's strategy space into binary options that depend on the buyer's order, thereby effectively addressing Question \ref{q2}.

With increasing concerns over data analysis-based discrimination in markets (e.g., \cite{mattioli_2012_on,a2021_zalescom,gunderson2012amazon}), our research further extends to exploring the impact of buyer order information protection. In ML model trading, the buyer could risk the leakage of order information, enabling the seller's discriminatory deception strategy. Considering potential regulations (such as the Robinson-Patman Act \cite{hovenkamp2000robinson} and Data Governance Act \cite{Specht-Riemenschneider2025-pq}) that aim to prevent a ``personalized'' trading strategy from the seller, we delve into our third key question:

\begin{question}
\label{q3}
How does protecting the buyer's order information affect the strategies and payoffs of the buyer and seller? 
\end{question}

The order information protection further complicates strategic interaction in model trading. Particularly, the seller needs to anticipate a range of possible buyer orders and verification strategies when making the model delivery decision, leading us to study seller and buyer reaction correspondence \cite{mas1995microeconomic}.

The key results and contributions of the paper are as follows:
\begin{itemize}
    \item \emph{A Novel Problem Formulation:} To the best of our knowledge, this is the first study on ML model trading under information asymmetry, where the self-interested seller could misrepresent model quality for payoff maximization. Additionally, we explore the impact of buyer order information protection in model markets, preventing the seller's potential discriminatory trading strategy.
    \item \emph{Effectiveness of Model Verification:} We enhance trading analysis by removing dominated strategies and integrating learning curves' power-law feature, enabling determination of equilibrium in model trading without order information protection. Our work reveals a pivotal finding: model verification mutually advantages buyers and sellers by diminishing information asymmetry. Specifically, cost-effective verification alleviates the strategic tension from information asymmetry, improving both parties' payoff.
    \item \emph{Impact of Order Information Protection:} Our exploration of order information protection in model trading indicates that such protection does not improve the payoff for either the buyer or seller. Notably, although protecting order information prevents discriminatory trading strategies from the seller, it also discourages the buyer from verifying the delivered model. This absence of model verification, in turn, motivates the seller to adopt model deception, resulting in less payoff for both the buyer and seller than that without order information protection.
    \item \emph{Scheme Design for Model Pricing:} We propose optimal pricing schemes for seller payoff maximization in model trading with and without order information protection. In particular, considering heterogeneous buyers in ML model markets, we design a pricing algorithm that adapts to the diversity of buyer model utility. Extensive numerical experiments demonstrate that, under information asymmetry, our pricing schemes achieve seller payoff close to the benchmark of complete information.
\end{itemize}

\section{Related Work}
\label{RW}
In this section, we review related works on both data markets and ML model markets.

Research in data markets, as summarized in surveys \cite{liang2018survey,Az2022,zhang2024surveydatamarkets}, has made impressive progress, particularly in privacy protection and data pricing (e.g., \cite{ghosh2011selling,Hynes2018-qr,muschalle2012pricing,cummings2015accuracy,Niu2018-hm,yu2017data,azcoitia2020try,koutris2012querymarket,koutris2013toward,koutris2015query,Cong2022-pf,az_icde}). Specifically, Ghosh \emph{et al.} in \cite{ghosh2011selling} studied data markets through the lens of differential privacy. Hynes \emph{et al.} in \cite{Hynes2018-qr} developed decentralized data markets with privacy-preserving smart contracts. Niu \emph{et al.} in \cite{Niu2018-hm} formulated a cost-based data pricing framework to compensate data owners for privacy losses. Koutris \emph{et al.} in \cite{koutris2012querymarket,koutris2013toward,koutris2015query} proposed the framework of a query-based data market, enabling flexible data evaluation and pricing. \emph{In contrast, this paper concentrates on the data-driven model market, which is relatively under-explored and more complex, given the unique characteristics of ML models.}

The study on the ML model market only emerged recently, as detailed in \cite{chen2019towards,agarwal2019marketplace,liu2020dealera,sunpeng}. Chen \emph{et al.} in \cite{chen2019towards} designed a model-based pricing framework, which directly prices ML models instead of raw data. Agarwal \emph{et al.} in \cite{agarwal2019marketplace} investigated a range of algorithms required for ML model markets. Liu \emph{et al.} in \cite{liu2020dealera} explored the effect of data usage restriction on the seller in model trading. Sun \emph{et al.} in \cite{sunpeng} presented a differential private federated learning model market. \emph{However, these works neglect the information asymmetry in model trading, where the seller could misrepresent model quality for payoff maximization.}

The remainder of this paper is organized as follows. Section \ref{SM} introduces the system model and presents the three-stage game formulation. Section \ref{S34} gives the equilibrium analysis in Stage 3 of the ML model trading process. Section \ref{S2} and Section \ref{S1} derive the buyer's optimal order decision in Stage 2 and the seller's optimal pricing scheme in Stage 1, respectively. Section \ref{OIP} extends our study to ML model trading with order information protection. Section \ref{S7} discusses pricing schemes under heterogeneous buyer utility. Section \ref{EXP} provides numerical results, and we conclude this paper in Section \ref{Con}. Due to the page limit, we leave detailed proof of results in the Appendix.

\section{System Model}
\label{SM}
This section introduces the ML model trading process without order information protection, and characterizes the decisions and payoffs of both the buyer and seller.

\subsection{ML Model Trading Process}
\label{SMPS}
We first briefly describe the ML model trading process, formulating it as the following three-stage sequential game:
\begin{itemize}
    \item \textbf{Stage 1:} The \emph{seller} publishes a set of ML models, each with a price and specific announced quality.
    \item \textbf{Stage 2:} The \emph{buyer} decides whether and which ML model to purchase from the set of published ML models. 
    \item \textbf{Stage 3:} If the buyer purchases an ML model, the \emph{seller} decides the delivery strategy, i.e., what true quality (that may be different from the announced quality in Stage 1) to deliver for the model ordered by the buyer. Meanwhile, the \emph{buyer} decides the verification strategy, i.e., whether to verify the delivered ML model with the local test data, which incurs associated costs. Additionally, the buyer sets the corresponding acceptance criterion (i.e., if verified, when to accept the delivered model).
\end{itemize}
\vspace{-8pt}\begin{figure}[ht]
    \centering
    \includegraphics[width=0.42\textwidth]{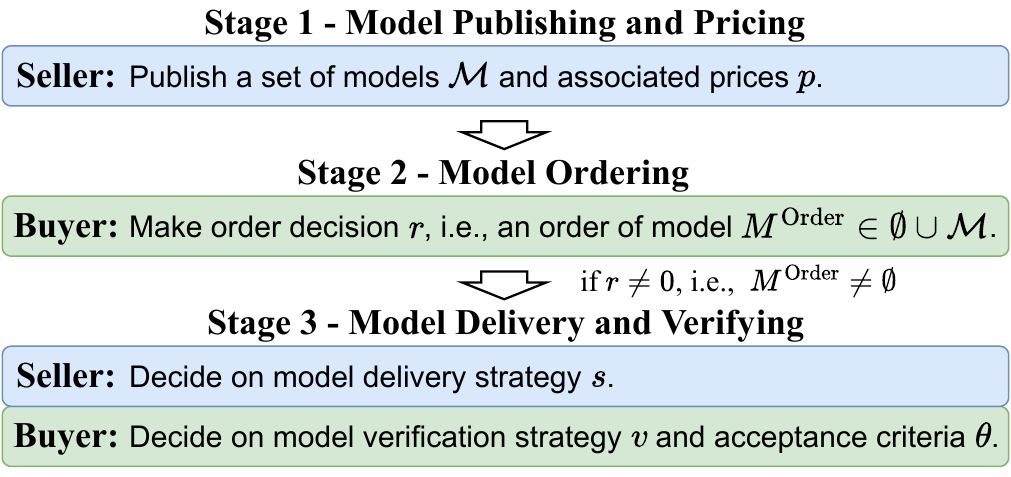}\vspace{-8pt}
    \caption{The ML Model Trading Process}\label{F1}
\end{figure}\vspace{-3pt}

Next, we elaborate on each stage in detail, as in Figure \ref{F1}.

\subsubsection{Stage 1 $-$ Model Publishing} In Stage 1, the seller trains a set of ML models with different quality levels and then publishes them with the corresponding prices:
\begin{itemize}
    \item \textit{Published Model Set:} Let $\mathcal{N}=\{1,2,\cdots,N\}$ denote the ML model index set, and let $\mathcal{M}=\{M_n, n\in\mathcal{N}\}$ denote the set of published models for sale. Following common conventions in the literature, e.g., \cite{agarwal2019marketplace,sizetest,Cong2022-pf}, we define model quality as the ratio of outcomes that meet the buyer's criteria, such as accurate predictions or losses below a predetermined threshold. Let $\alpha_n$ represent this quality metric of model $M_n$, where $\alpha_n$ is in the range $[0,1]$ for each $n\in\mathcal{N}$. Without loss of generality, we assume a larger index indicates a higher quality ML model, i.e., $\alpha_n\leq \alpha_m$ if $n\leq m$, for all $n,m\in\mathcal{N}$. The buyer's utility $U_n$ from employing model $M_n$ increases with model quality $\alpha_n$. Therefore, $0 \leq U_n \leq U_m$ whenever $n \leq m$, for all $n, m \in \mathcal{N}$.
    \item \textit{Model Prices:} We use $p_{n}$ to represent the published price of model $M_{n}$, for each $n\in\mathcal{N}$. Model prices align with quality: $0 \leq p_n \leq p_m$ when $n \leq m$ for all $n,m \in \mathbb{N}$.
\end{itemize}

\subsubsection{Stage 2 $-$ Model Ordering} After the seller publishes ML models in Stage 1, the buyer decides whether to purchase and (if so) which ML model $M^{\mathrm{Order}}$ to order from the model set $\mathcal M$. We denote the buyer's order decisions as $r\in\{0\}\cup\mathcal{N}$, including not ordering as an option. Hence,
\begin{equation}
    M^{\mathrm{Order}}=\left\{
        \begin{array}{ll}
        \emptyset, &\quad\textrm{if $r=0$},\\
        M_{r}, &\quad\textrm{if $r\in\mathcal{N}$}.
        \end{array}
    \right.
\end{equation}

\subsubsection{Stage 3 $-$ Model Delivery and Model Verifying} 
\begin{figure}[ht]\vspace{-12pt}
    \centering
    \includegraphics[width=0.3\textwidth]{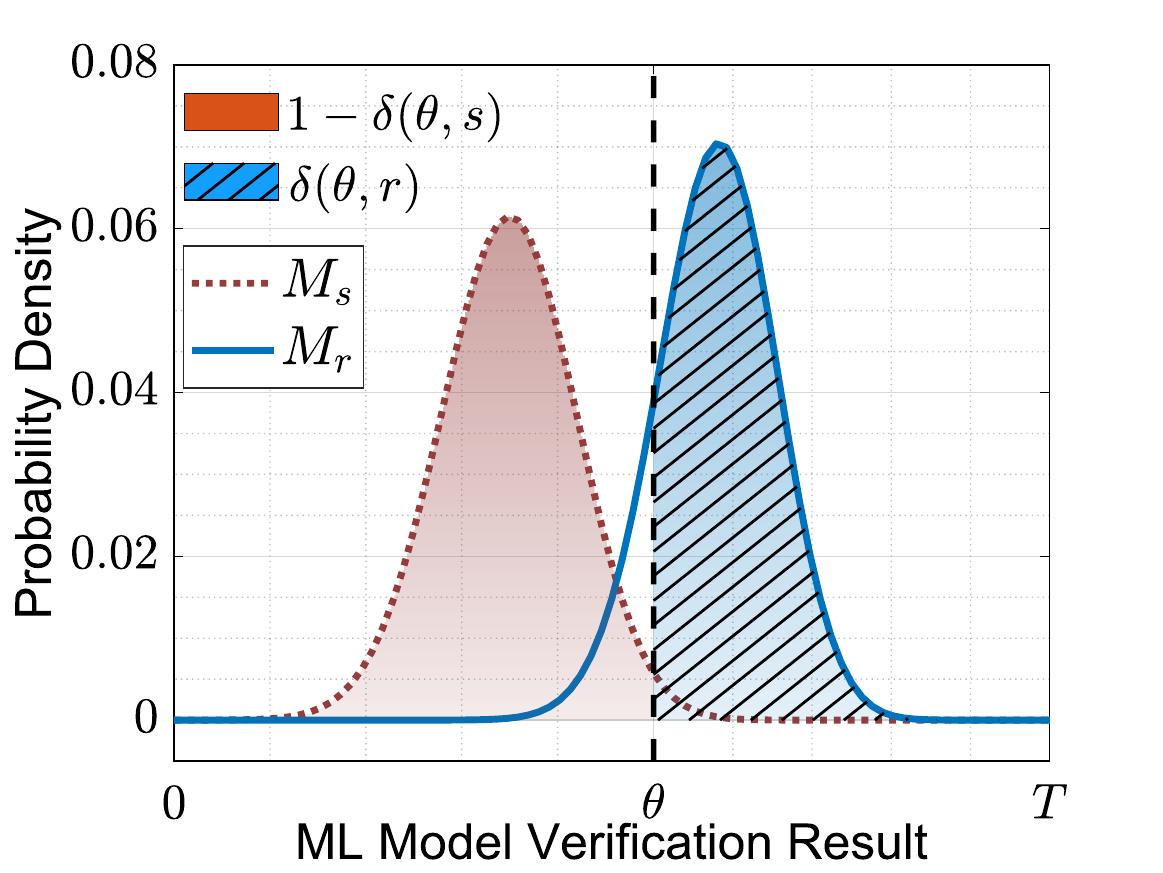}
    \vspace{-6pt}
    \caption{Illustration of Acceptance Criterion and Probability}\vspace{-8pt}
    \label{DIS}
\end{figure}

In Stage 3, the seller decides on the model delivery strategy, and the buyer decides on the verification strategy and acceptance criterion, as detailed below:

\begin{itemize}
    \item \textit{Model Delivery Strategy:} We define the seller's delivery strategy as $s\in\mathcal{N}$, indicating that the actual quality of the delivered ML model $M^{\mathrm{Deliver}}\in\mathcal{M}$, i.e., $M^{\mathrm{Deliver}}=M_s$.
    \item \textit{Model Verification Strategy:} Before paying the seller according to the price announced in Stage 1, the buyer can decide whether to verify the delivered ML model quality with his\footnote{For ease of discussion, we will use ``she'' to refer to the seller and ``he'' to refer to the buyer. This terminology choice does not imply any gender bias.} local test data. We denote the buyer's test data size as $T\in\mathbb{N^+}$ and use $v\in\{NV,V\}$ to represent his verification strategy:
\begin{equation}
    v=\left\{
        \begin{array}{ll}
             NV, &\textrm{not verify},\\
             V, &\textrm{verify}.
        \end{array}
    \right.
\end{equation}
The model verification, however, is often imperfect. For example, the performance of the same ML model can vary on different test data sets \cite{raschka2018model}, where the sampling bias of test data \cite{kramer2013importance} will affect verification results. 

\item \textit{Acceptance Criterion:} When adopting model verification, the buyer must establish an acceptance criterion to determine which verification results warrant accepting the delivered model, given that these outcomes are imperfect. We denote $\theta\in\mathbb{N}$ as the buyer's acceptance criterion, which is the least model quality demonstrated in verification results that the buyer will accept. For simplicity, we consider $\theta\in[0,T+1]$, where $\theta=0$ means the buyer accepts any model, regardless of verification outcomes, while $\theta=T+1$ is the opposite, i.e., reject all models. With criterion $\theta$ and delivery strategy $s\in\mathcal{N}$, we can determine the buyer's acceptance probability $\delta(\theta,s)$ of the delivered model $M^{\mathrm{Deliver}}=M_s$. Specifically, under criterion $\theta$, when the buyer receives a model $M_s$ with quality metric $\alpha_s$, we calculate the acceptance probability $\delta(\theta,s)$ in \eqref{binocal}, which falls between $0$ and $1$.
\begin{equation}\label{binocal}
    \delta(\theta,s)=\sum\nolimits_{i=\theta}^T \binom{T}{i}\alpha_s^i(1-\alpha_s)^{T-i}.
\end{equation}
From \eqref{binocal}, given $T$ test data samples, $\delta(\theta,s)$ is the cumulative probability that model $M_s$ meets the buyer's criterion $\theta$ across all possible verification outcomes. Figure \ref{DIS} provides a visual representation of the relationships between acceptance criterion, probability, and ML model quality. The shaded blue area with diagonal lines indicates the buyer's acceptance probability $\delta(\theta,r)$ under criterion $\theta$ when the delivered model $M^{\mathrm{Deliver}}=M_r$.
\end{itemize}

\subsection{Expected Buyer Payoff}
The buyer payoff includes the expected utility of employing the ML model, expected payment for the model, and the verification cost. Next, we introduce each part in more detail.

\subsubsection{Expected Model Utility} Model utility depends on the actual quality of the delivered ML model. Note $U_{n}$ is the buyer's utility of employing model $M_{n}$, with $0\leq U_{n}\leq U_{m}$ if $n\leq m$, for $n,m\in\mathcal{N}$. However, as the model quality may not be what the seller claims, the buyer's expected utility, when he orders model $M_r$ while the seller delivers a model $M_s$, is:
\begin{subnumcases}{\label{UM}\mathbb{E}[U(r,s,v,\theta)]=}
0,&\textrm{if $r=0$},\label{5a}\\
U_{s},&\textrm{if $r\in\mathcal{N}$, $v=NV$},\label{5b}\\
\delta(\theta,s)\cdot U_{s},&\textrm{if $r\in\mathcal{N}$, $v=V$}.\label{5c}
\end{subnumcases}

From \eqref{5a}, the utility is zero when the buyer does not order any model. If the buyer orders model $M_r$ without verification ($v=NV$), as in \eqref{5b}, he will directly accept it and the utility will be $U_s$, which depends on delivery strategy $s$. Finally, in \eqref{5c}, if the buyer orders $M_r$ and verifies the delivered model before actual payment, the verification will succeed (delivered model $M_s$ is tested with a quality no less than the buyer's acceptance criterion $\theta$) with a probability $\delta(\theta,s)$, and thus the expected utility is $\delta(\theta,s)U_{s}$.

\subsubsection{Expected Model Payment}
The expected payment for the ML model depends on the trading decisions as follows:
\begin{subnumcases}{\label{PM}\mathbb{E}[P(r,s,v,\theta)]=}
0, &\textrm{if $r=0$},\label{6a}\\
p_{r}, &\textrm{if $r\in\mathcal{N}$, $v=NV$},\label{6b}\\
\delta(\theta,s)\cdot p_{r}, &\textrm{if $r\in\mathcal{N}$, $v=V$}\label{6c}.
\end{subnumcases}

From \eqref{6a}, the payment is zero when the buyer does not order any model from the seller. If the buyer orders an ML model $M_r$ without verification, he will pay $p_r$, as stated in \eqref{6b}. If the buyer decides to verify the delivered model, the payment depends on the verification result with an acceptance probability $\delta(\theta,s)$, as in \eqref{6c}.

\subsubsection{Verification Cost} For simplicity, we assume the buyer incurs a cost $C_T>0$ to collect test data for model verification. This allows for flexibility in capturing different relationships between verification cost $C_T$ and the test sample size $T$.

\subsubsection{Expected Buyer Payoff} In summary, the expected buyer payoff $\pi_{buyer}(r,s,v,\theta)$ in the ML model trading is:
\begin{align}
\label{Payoff}
    \pi_{buyer}(r,s,v,\theta)=&\mathbb{E}[U(r,s,v,\theta)-P(r,s,v,\theta)]\nonumber\\
    &-C_T\cdot\boldsymbol{1}_{v=V},
\end{align}
where $\mathbb{E}[U(r,s,v,\theta)]$ and $\mathbb{E}[P(r,s,v,\theta)]$ are defined in \eqref{UM} and \eqref{PM}, respectively. The indicator $\boldsymbol{1}_{v=V}$ indicates that the buyer will verify the delivered model. The buyer aims to make optimal decisions on $r$, $v$ and $\theta$ to maximize $\pi_{buyer}(r,s,v,\theta)$.

\subsection{Expected Seller Payoff}
The seller's payoff not only includes the buyer’s expected payment (seller’s income) but also the expected cost of selling an ML model, which we will discuss next.

\subsubsection{Expected Model Cost} For the seller, ML model trading incurs the cost of collecting training data from data owners and the cost of processing the data. Since model is a digital asset derived from data, we focus on the cost of compensating data owners' sensitive privacy loss, as illustrated in \cite{liu2020dealera}. That is, we spread all one-time costs\footnote{As a digital product, the replication cost of ML models is negligible. Thus, the marginal cost of ML models caused by one-time cost is insignificant compared to privacy compensation in each ML model trading.} (e.g., computation cost) across each trading, and normalize them to zero. Let $C_n$ denote the cost of trading model $M_n$, with $0<C_n<C_m$ if $n<m$, for each $ n\in\mathcal{N}$ (i.e., model quality improvement often requires a larger training data size and hence a higher cost). We summarize the expected cost of selling a model as follows:
\begin{subnumcases}{\label{CM}\mathbb{E}[C_M(r,s,v,\theta)]=}
\textrm{$0$},&\hspace{-8.5pt}\textrm{if $r=0$},\\
\textrm{$C_{s}$},&\hspace{-8.5pt}\textrm{if $r\in\mathcal{N}$, $v=NV$},\\
\textrm{$\delta(\theta,s)\cdot C_{s}$},&\hspace{-8.5pt}\textrm{if $r\in\mathcal{N}$, $v=V$}.
\end{subnumcases}

Note that privacy compensation occurs when the buyer applies the ML model in practice, i.e., the seller incurs model cost when the buyer accepts the delivered model. Hence, the explanation of \eqref{CM} is similar to that of \eqref{UM}, which also depends on whether the buyer accepts the delivered ML model.

\subsubsection{Expected Seller Payoff} To summarize, the expected seller payoff is the difference between income and model cost:
\begin{equation}
\label{Revenue} \pi_{seller}(r,s,v,\theta)=\mathbb{E}[P(r,s,v,\theta)-C_M(r,s,v,\theta)]. 
\end{equation}

Next, we analyze the three-stage sequential game and derive the trading equilibrium through backward induction \cite{mas1995microeconomic}.

\section{Model Delivery and Verifying in Stage 3}
\label{S34}
In this section, we first formulate the simultaneous game between the buyer and seller in Stage 3 of the model trading process, and then derive the Nash equilibrium of this game.

\subsection{Trading Game Formulation}
Given the seller's published model information in Stage 1 and the buyer's order decision\footnote{Note that Stage 3 only exists after the buyer orders a model, i.e., $r\neq0$.} $r$ in Stage 2, we formulate the trading game in Stage 3 as follows:
\begin{game}[Trading Game in Stage 3]
\label{Game1}
The trading game in Stages 3 is as follows:
\begin{itemize}
    \item Players: The seller and the buyer.
    \item Strategies: The seller decides the delivery strategy $s\in\mathcal{N}$. The buyer decides the verification strategy $v\in\{NV,V\}$ and acceptance criterion $\theta\in[0,T+1]\cup\{NA\}$, where $NA$ stands for not applicable if $v=NV$.
    \item Payoffs: The seller maximizes her expected payoff in \eqref{Revenue}. The buyer maximizes his expected payoff in \eqref{Payoff}.
\end{itemize}
\end{game}
\subsection{Equilibrium of the Trading Game \ref{Game1}}
\label{DVE34}

In Trading Game \ref{Game1}, the seller and buyer decide strategies to maximize their respective payoffs. Next, we will derive the Nash equilibrium, in which no player can increase individual payoff by deviating unilaterally. We first focus on pure strategy Nash equilibrium (PNE), where each player chooses only one strategy at equilibrium, as introduced in Definition \ref{DPNE}.

\begin{definition}[Pure Strategy Nash Equilibrium of Game \ref{Game1}]\label{DPNE}
A pure strategy profile ($s^*$,($v^*,\theta^*$)) is a PNE of Game \ref{Game1} if both the seller and buyer's decisions maximize their payoffs given each other's equilibrium strategy, i.e.,
\begin{subequations}
    \begin{align}
    \pi_{seller}(s^*,(v^*,\theta^*))&\geq \pi_{seller}(s,(v^*,\theta^*)),\\
    \pi_{buyer}(s^*,(v^*,\theta^*))&\geq \pi_{buyer}(s^*,(v,\theta)),
\end{align}
\end{subequations}
for each $s\in\mathcal{N}$, $v\in\{NV,V\}$  and $\theta\in[0,T+1]\cup\{NA\}$.
\end{definition}

Depending on the buyer's order decision $r$ in Stage 2, we analyze the equilibrium in two cases.

\subsubsection{The buyer orders $M_1$ in Stage 2} When the buyer orders the lowest quality model ($r=1$) in Stage 2, he does not need to verify it. The reason is that any delivered model $M_s$ in set $\mathcal{M}$ will satisfy the quality requirement of the buyer's order, i.e., $\alpha_1\leq\alpha_s$, for any $s\in\mathcal{N}$. Meanwhile, the seller cannot improve her payoff by delivering the buyer a model better than $M_1$, leading to a unique PNE in Proposition \ref{rL}.
\begin{proposition}\label{rL}
Given $r=1$ in Stage 2, the unique PNE of Game \ref{Game1} is $(s^*,(v^*,\theta^*))=(1, (NV,NA))$.
\end{proposition}
We give the detailed proof of Proposition \ref{rL} in the Appendix.
\subsubsection{The buyer orders $M_r$ ($r\in\mathcal{N}\setminus\{1\}$) in Stage 2} 
\begin{figure}[ht]
    \centering
    \includegraphics[width=0.3\textwidth]{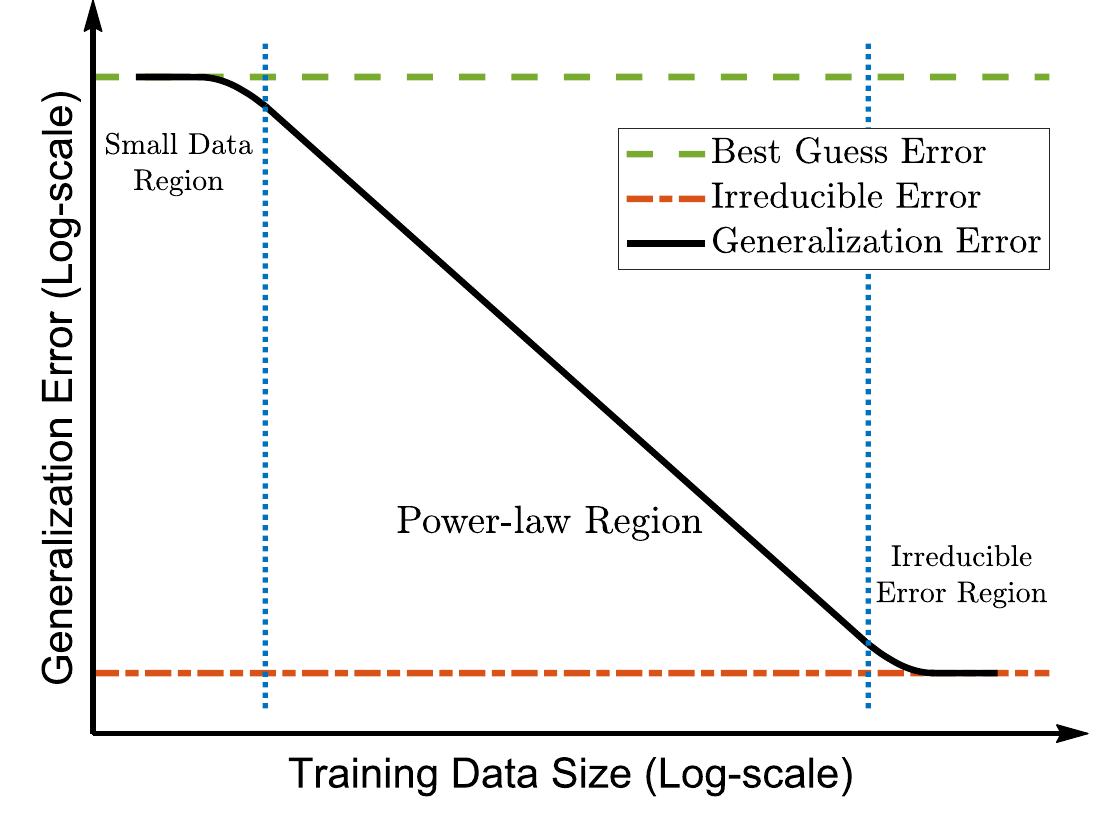}
    \vspace{-13pt}
    \caption{Sketch of Power-law Learning Curves}
    \label{powerlaw}\vspace{-5pt}
\end{figure}

When the buyer orders a model better than $M_1$, i.e., $r>1$, the equilibrium analysis involves deriving conditions under which it is economically and effectively beneficial for the buyer to verify the model. To facilitate this, we begin with the fundamental relationship\footnote{With very small data sizes, the ML model performs as well as random guessing, achieving the best guess generalization error. In contrast, with very large data sizes, the model reaches irreducible generalization error, i.e., the information-theoretic error lower bound \cite{hestness2017deep}, as illustrated in Figure \ref{powerlaw}.} between ML model quality (quantified by generalization error, which measures the model's performance on unseen data) and training data size, shown in Figure \ref{powerlaw}. The power-law nature of learning curves, as established in \cite{hestness2017deep,amari1993universal,amari1992four}, indicates that achieving each additional unit of model quality improvement requires logarithmically increasing amounts of training data. This relationship directly influences the cost-quality trade-off in model development.
\vspace{-8pt}

Using the learning curves' power-law feature, we apply the method of \emph{iterated elimination of strictly dominated strategies}, leading to Proposition \ref{PA}.
\begin{proposition}
\label{PA}
Given $r\in\mathcal{N}\setminus\{1\}$ in Stage 2, the seller will only deliver either $M_r$ or $M_1$ at the equilibrium of Game \ref{Game1}.
\end{proposition}

We give the detailed proof of Proposition \ref{PA} in the Appendix. With Proposition \ref{PA}, we then delve into Nash equilibrium of ordering $M_r$ ($r\in\mathcal{N}\setminus\{1\}$). This analysis considers both cases where $p_r\leq U_r$ and $p_r > U_r$, though we focus on the former case as it leads to more interesting strategic interactions. In the latter case, since the seller never delivers a model with quality higher than what the buyer pays for, the buyer would receive negative payoff and thus would not purchase the model.


\textit{a) Uneconomical Model Verification:} When verification is uneconomical against model deception, i.e., $C_T>p_r-U_1$, there is a unique PNE, as illustrated in Proposition \ref{rHu}.

\begin{proposition}\label{rHu}
Consider $r\in\mathcal{N}\setminus\{1\}$ in Stage 2 with $p_r\leq U_r$. When verification is uneconomical, i.e., $C_T>p_r-U_1$, the unique PNE of Game \ref{Game1} is $(s^*,(v^*,\theta^*))=(1,( NV,NA))$.
\end{proposition}

We give the detailed proof of Proposition \ref{rHu} in the Appendix. The uneconomical condition in Proposition \ref{rHu} implies that the cost of model verification is higher than its expected gain, where $p_r-U_1$ characterizes the highest payoff loss the buyer's verification can save. Therefore, the buyer will not verify the delivered model, and the seller will invariably deliver the lowest quality model, leading to persistent model deception.

\textit{b) Economical but Ineffective Model Verification:} When model verification is economically feasible but fails to prevent the seller's model deception, we summarize the corresponding unique PNE in Proposition \ref{rHi}.

\begin{proposition}\label{rHi}
    Consider $r\in\mathcal{N}\setminus\{1\}$ in Stage 2 with $p_r\leq U_r$. When verification is economical but ineffective, i.e.,
    \begin{itemize}
        \item $C_T<p_r-U_1$, and
        \item $\left(p_{r}-C_{r}\right)\cdot\delta(\theta,r)\leq\left(p_{r}-C_{1}\right)\cdot\delta(\theta,1)$, for any $\theta\in\{\theta_r\in[0,T+1]:C_T<(p_r-U_1)\cdot(1-\delta(\theta_r,1))\}$,
    \end{itemize}
    the unique PNE of Game \ref{Game1} is $(s^*,(v^*,\theta^*))=(1, (V,T+1))$.
\end{proposition}

We give the detailed proof of Proposition \ref{rHi} in the Appendix. From Proposition \ref{rHi}, with economical but ineffective model verification, the seller always performs deception by delivering the lowest quality model $M_1$, as this yields the highest payoff, even when considering the probabilistic verification result. In response, the buyer minimizes the trading loss by rejecting the delivered model upon verification, i.e., setting $\theta=T+1$.

\textit{c) Economical and Effective Model Verification:} When model verification is economical and effective against model deception, we show that there does not exist a PNE. If the buyer does not verify the model, the seller will opt for model deception. Conversely, if the buyer verifies, the seller will choose truthful delivery. Therefore, in pure strategies, a mutual best response is non-existent. This leads us to consider the mixed strategy Nash equilibrium (MNE), allowing players to probabilistically select multiple strategies at equilibrium. We first present the mixed strategy of Game \ref{Game1} in Definition \ref{DMS}, followed by the definition of MNE in Definition \ref{DMSNE}.

\begin{definition}[Mixed Strategy of Seller and Buyer]\label{DMS}
The mixed strategies of seller and buyer are vectors $\boldsymbol{\phi}_s \in \Phi_S$, $\boldsymbol{\phi}_v \in \Phi_V$:
\begin{subequations}
        \begin{align}
             \boldsymbol{\phi}_s&\triangleq\left(\textrm{Prob}[s=n], \forall n\in\mathcal{N}\right),\\
             \boldsymbol{\phi}_v&\triangleq\left(\textrm{Prob}[v=NV],\textrm{Prob}[v=V]\right),
        \end{align}
\end{subequations}
where $\Phi_S$ and $\Phi_V$ are the simplex over set $\mathcal{N}$ and $\{NV,V\}$, i.e., $\sum_{s\in\mathcal{N}}\textrm{Prob}[s]=1$ and $\sum_{v\in\{NV,V\}}\textrm{Prob}[v]=1$.
\end{definition}

\begin{definition}[Mixed Strategy Nash Equilibrium of Game \ref{Game1}]
\label{DMSNE}
 A mixed strategy profile ($\boldsymbol{\phi}_s^*$,($\boldsymbol{\phi}_v^*,\theta^*$)) is an MNE of Game \ref{Game1} if both the seller and buyer's decisions maximize their payoffs given each other's equilibrium strategy, i.e., 
 \begin{subequations}
     \begin{align}
    \pi_{seller}(\boldsymbol{\phi}_s^*,(\boldsymbol{\phi}_v^*,\theta^*))&\geq \pi_{seller}(\boldsymbol{\phi}_s,(\boldsymbol{\phi}_v^*,\theta^*))),\\
    \pi_{buyer}(\boldsymbol{\phi}_s^*,(\boldsymbol{\phi}_v^*,\theta^*)))&\geq \pi_{buyer}(\boldsymbol{\phi}_s^*,(\boldsymbol{\phi}_v,\theta)),
\end{align}
 \end{subequations}
for each $\boldsymbol{\phi}_v\in\Phi_V$, $\boldsymbol{\phi}_s\in\Phi_S$ and $\theta\in[0,T+1]\cup\{NA\}$.
\end{definition}
Next, we derive the MNE of Game \ref{Game1} when model verification is economical and effective. We determine the necessary condition $\Gamma_r$ in \eqref{Gma} for the equilibrium acceptance criterion:
\begin{subnumcases}{\Gamma_r(\theta):\label{Gma}}
&\hspace{-6pt}$C_T<(p_r-U_1)\cdot(1-\delta(\theta,1))$,\\
&\hspace{-6pt}$\left(p_{r}-C_{1}\right)\cdot\delta(\theta,1)<\left(p_{r}-C_{r}\right)\cdot\delta(\theta,r)$,\\
&\hspace{-6pt}$\sum\limits_{i=0}^{\theta-1}\binom{T}{i}\left((\frac{\alpha_r}{1-\alpha_r})^{-\theta+i}-(\frac{\alpha_1}{1-\alpha_1})^{-\theta+i}\right)/C_T$\nonumber\\
&\hspace{-6pt}$+\frac{\alpha_r^{-\theta}(1-\alpha_r)^{\theta-1}}{U_r-p_r}+\frac{\alpha_1^{-\theta}(1-\alpha_1)^{\theta-1}}{p_r-U_1}\leq 0$.
\end{subnumcases}
Using $\Gamma_r$, we establish buyer's equilibrium criterion $\theta^*(r)$ as
\begin{equation}\label{tstar}
    \theta^*(r)=\min\{\theta\in[0,T+1]:\Gamma_r(\theta)\},
\end{equation}
and derive the equilibrium probability distribution over model delivery and verification strategy in Proposition \ref{NE DV}. For simplicity, we denote the equilibrium acceptance probability $\delta(\theta^*(r),1)$ and $\delta(\theta^*(r),r)$ as $\delta_{r,1}^*$ and $\delta_{r,r}^*$, respectively.
\begin{proposition}
\label{NE DV}
Consider $r\in\mathcal{N}\setminus\{1\}$ in Stage 2 with $p_r\leq U_r$. When verification is economical and effective, i.e.,
\begin{subnumcases}{\label{PEEC}}
&$\hspace{-8pt}C_T<(p_r-U_1)\cdot(1-\delta(\theta,1))$, and\\
&$\hspace{-8pt}\left(p_{r}-C_{1}\right)\cdot\delta(\theta,1)<\left(p_{r}-C_{r}\right)\cdot\delta(\theta,r)$,
\end{subnumcases}
where $\theta\in[0,T+1]$, the unique MNE of Game \ref{Game1} is: 
\begin{itemize}
    \item Seller: delivers $M_1$ with probability $\textrm{Prob}{[s^*=1]}$ in \eqref{9d} and delivers $M_r$ with probability $1-\textrm{Prob}{[s^*=1]}$. 
    \item Buyer: sets $\theta^*(r)=\min\{\theta\in[0,T+1]:\Gamma_r(\theta)\}$ in \eqref{Gma}, verifies with probability $\textrm{Prob}{[v^*=V]}$ in \eqref{9b} and does not verify with probability $1-\textrm{Prob}{[v^*=V]}$.
\end{itemize}
\begin{subnumcases}{\label{ES}}
&$\hspace{-8pt}\textrm{Prob}{[s^*=1]} = \frac{(1-\delta_{r,r}^*)(U_{r}-p_{r})+C_T}{(1-\delta_{r,1}^*)(p_{r}-U_{1})+(1-\delta_{r,r}^*)(U_{r}-p_{r})}$,\label{9d}\\
&$\hspace{-8pt}\textrm{Prob}{[v^*=V]}=\frac{C_{r}-C_{1}}{(1-\delta_{r,1}^*)(p_{r}-C_{1})-(1-\delta_{r,r}^*)(p_{r}-C_{r})}$\label{9b}.
\end{subnumcases}
\end{proposition}

We give the detailed proof of Proposition \ref{NE DV} in the Appendix. Proposition \ref{NE DV} indicates that a rise in verification cost $C_T$ leads to a higher equilibrium acceptance criterion $\theta^*(r)$, making the buyer less inclined to accept the verified model. This stems from the increased loss the buyer incurs, attributed to the higher verification cost, necessitating a more stringent acceptance criterion to prevent the adoption of inferior models. Yet, this stricter criterion, alongside the rising verification cost, may incentivize the seller to engage in model deception.

\section{The Optimal Order Decision in Stage 2}
\label{S2}
After investigating the equilibrium in Stage 3, this section studies the buyer's optimal order decision in Stage 2.

\subsection{Buyer Payoff Optimization Problem}
In Stage 2, the buyer aims to make the optimal order decision $r^*$ to maximize his payoff in \eqref{Payoff}. To emphasize the dependence of Stage 3 decisions on order decision $r$, we write the equilibrium of Game \ref{Game1} as $(s^*(r),(v^*(r),\theta^*(r)))$. Then, we formulate the buyer's payoff optimization problem as follows:

\begin{problem}\label{P1}
(Buyer Payoff Optimization Problem).
\begin{subequations}
\begin{align}
\max\quad&\mathbb{E}[U(r,s^*(r),v^*(r),\theta^*(r))-P(r,s^*(r),v^*(r),\theta^*(r))]\nonumber\\
&-C_T\cdot\boldsymbol{1}_{v^*(r)=V},\\
var.\quad&r\in \{0\}\cup\mathcal{N}.
\end{align}
\end{subequations}
\end{problem}
\subsection{The Optimal Order Decision}
\label{S5}
Based on the equilibrium investigation in Stage 3, we can derive the buyer's optimal order decision $r^*$ in Stage 2, along with associated conditions. Specifically, we divide the discussion regarding $r^*$ into three cases, as detailed below.
\subsubsection{$r^*\in\mathcal{N}\setminus\{1\}$ in Stage 2} \label{B1}
We clarify the conditions under which the optimal order decision $r^*$ is any choice other than the lowest quality model $M_1$. Per Nash equilibrium insights from Propositions \ref{rHu}-\ref{NE DV}, uneconomical or ineffective verification prompts the seller to consistently provide model $M_1$, resulting in the buyer's overpayment. Thus, it is essential to satisfy the economical and effective verification in \eqref{PEEC} for the decision to order model $M_n$, where $n\in\mathcal{N}\setminus\{1\}$ and is determined by the following condition $\Lambda_n^{EE}$:
\begin{subnumcases}{\Lambda^{EE}_n:\label{CEE}}
C_T<(1-\delta^*_{n,1})(p_n-U_1),\label{17a}\\
\delta_{n,1}^*(p_{n}-C_{1})<\delta_{n,n}^*(p_{n}-C_{n}).
\end{subnumcases}

Under condition $\Lambda_n^{EE}$, the buyer's payoff of ordering $M_n$ is $U_n-p_n-\Delta\pi_n$, where $\Delta\pi_n$ in \eqref{(9)} indicates the payoff reduction from information asymmetry:
\begin{equation}
    \label{(9)} \Delta\pi_n=\frac{(U_{n}-U_{1})C_T+(1-\delta^*_{n,n})(U_{n}-U_{1})(U_{n}-p_{n})}{(1-\delta^*_{n,1})(p_{n}-U_{1})+(1-\delta^*_{n,n})(U_{n}-p_{n})}.
\end{equation}

To arrive at optimal decision $r^*=n\in\mathcal{N}\setminus\{1\}$, model $M_{n}$ must bring buyer the highest payoff among all other decisions. Overall, we summarize conditions for ordering $M_n$ as $\Lambda_n$:
\begin{subnumcases}{\Lambda_n:\label{CH}}
\Lambda^{EE}_n,\\
\max\{0,U_1-p_1\}\leq U_n-p_n-\Delta\pi_n,\\
n=\argmax\nolimits_{m\in\mathcal{N}\setminus\{1\}}\textrm{ }U_m-p_m-\Delta\pi_m.\label{13c}
\end{subnumcases}
\subsubsection{$r^*=1$ in Stage 2} We examine when the buyer's maximum payoff results from ordering the lowest quality model $M_1$, i.e., $r^*=1$. As detailed in Section \ref{B1}, $r^*$ equals $1$ if and only if $U_1-p_1$ is non-negative and no other condition $\Lambda_n$ is met for any $n\in\mathcal{N}\setminus\{1\}$. For ease of exposition, we denote such conditions as $\Lambda_1:\{\neg\Lambda_n,\forall n\in\mathcal{N}\setminus\{1\}\}\land\{0\leq U_1-p_1\}$.

\subsubsection{$r^*=0$ in Stage 2} When the payoff of ordering any model is negative, the optimal order decision is not to order.

To summarize the discussions of three possible decisions above, Theorem \ref{OD} concludes the buyer's optimal order decision $r^*$ in Stage 2 of the ML model trading.

\begin{thm}
\label{OD}
The buyer's optimal order decision $r^*$ is:
\begin{subnumcases}{\label{TH1 E} r^*=}
n\in\mathcal{N}, &\textrm{if $\Lambda_n$ holds},\\
0, &\textrm{otherwise}.
\end{subnumcases}
\end{thm}
We give the detailed proof of Theorem \ref{OD} in the Appendix. Theorem \ref{OD} identifies the key condition $\Lambda_n^{EE}$ outlined in \eqref{CEE} for economical and effective verification, enabling buyers to sidestep the trading trap of the seller. 

Example \ref{ExU} demonstrates this with a case involving two ML models for sale, i.e., $N=2$.
\begin{example}[Trading Trap]
\label{ExU}
With two models for sale, where $U_1=11$ and $U_2=1000$, and corresponding prices $p_1=9$ and $p_2=10$, a negligible verification cost of $C_T=0.0001$ leads the buyer to optimally choose $r^*=1$. This choice yields the maximum payoff of $U_1-p_1=2$.
\end{example}

Example \ref{ExU} seems counter-intuitive, as $U_2-p_2=990$ is much higher than $U_1-p_1=2$, i.e., high-quality model $M_2$ is much more attractive for the buyer, and $C_T$ is negligible. However, if the seller offers low-quality model $M_1$ when the order is $r=2$, the buyer will still accept it and pay the price $p_2$, since $U_1-p_2=1>0$. That is, the buyer will never verify the delivered model, as doing so will not improve his payoff (i.e., uneconomical verification). Model verification thus becomes an empty threat that the buyer never implements, based on which the seller always conducts model deception by delivering $M_1$. To avoid this trap, Theorem \ref{OD} introduces condition $\Lambda_n^{EE}$ in \eqref{CEE}, which ensures an economical and effective verification, thereby maximizing buyer payoff.

\subsection{The Optimal Buyer Payoff}
\label{5.3}
Given the optimal order decision $r^*$ in Theorem \ref{OD}, we can compute the resultant optimal buyer payoff $\pi_{buyer}^*$ as follows:
\begin{subnumcases}{\label{(14)}\pi_{buyer}^*=}
0, &\textrm{if $r^*=0$},\\
U_{1}-p_{1}, &\textrm{if $r^*=1$},\\
U_{r^*}-p_{r^*}-\Delta\pi_{r^*}, &\textrm{if $r^*\in\mathcal{N}\setminus\{1\}$}.
\end{subnumcases}

To investigate the effectiveness of model verification in tackling information asymmetry, we compare the buyer's optimal payoff $\pi_{buyer}^*$ to that of \textit{complete information} benchmark. In particular, with complete information, the ML model quality is transparent to both the seller and buyer, which implies that the buyer's payoff from purchasing an ML model is the difference between model utility and price, as expressed in \eqref{benchmarkbuyer}.
\begin{subnumcases}{\label{benchmarkbuyer}\pi^{benchmark}_{buyer}=}
0, &\textrm{if $r^*=0$},\\
U_{r^*}-p_{r^*}, &\textrm{if $r^*\in\mathcal{N}$}.
\end{subnumcases}

By comparing the optimal payoffs in \eqref{(14)} and \eqref{benchmarkbuyer}, although the buyer can adopt verification in ML model trading, information asymmetry still reduces his optimal payoff by $\Delta\pi_{r^*}$ in \eqref{(9)}. This payoff reduction is due to the following two reasons:
\begin{itemize}
    \item \textit{Verification Cost:} From Proposition \ref{NE DV}, when purchasing $M_{r^*}$ where $r^*\in\mathcal{N}\setminus\{1\}$, the buyer will adopt a probabilistic verification strategy against the seller's potential model deception. The cost of model verification thus decreases the buyer's optimal payoff.
    \item \textit{Imperfect Verification:} With imperfect model verification, the buyer may overestimate or underestimate the delivered model quality, thereby reducing the buyer's optimal payoff compared to the \textit{complete information} benchmark.
\end{itemize}


\section{The Optimal Pricing Scheme in Stage 1}
\label{S1}
Following the analysis of Nash equilibrium in Stage 3 and the buyer's optimal order decision in Stage 2, we proceed to determine the seller's optimal pricing scheme in Stage 1.

\subsection{Seller Payoff Optimization Problem}
\label{6.1}
In Stage 1, the seller decides on a model pricing scheme $\boldsymbol{p}=\{p_n,\forall n\in\mathcal{N}\}$ to maximize payoff $\pi_{seller}$ in \eqref{Revenue}. Considering the dependence of the buyer's optimal order decision $r^*$ on model prices $\boldsymbol{p}$, we represent it as $r^*(\boldsymbol{p})$. This allows us to formulate the seller's payoff optimization problem as Problem \ref{P2}, which contains a sub-problem regarding buyer payoff optimization in Problem \ref{P1}.

\begin{problem}\label{P2}
Seller Payoff Optimization Problem.
\begin{subequations}
\begin{align}
\max&\quad \mathbb{E}[P(r^*(\boldsymbol{p}))-C_M(r^*(\boldsymbol{p}))],\\
s.t.&\quad  r^*=\argmax_{r\in\{0\}\cup\mathcal{N}}\pi_{buyer}(r,\boldsymbol{p}),\\
&\quad  0\leq p_i\leq p_j,\quad\forall i\leq j,\textrm{ }\forall i,j\in\mathcal{N},\\
var.&\quad \boldsymbol{p}=\{p_n,\forall n\in\mathcal{N}\}.
\end{align}
\end{subequations}
\end{problem}

\subsection{The Optimal Pricing Scheme}
\label{6.2}
To address Problem \ref{P2}, we begin by discussing the upper bound of each ML model price, with which we then determine the optimal pricing scheme $\boldsymbol{p}^*$ and associated conditions.

\subsubsection{Upper Bound of Model Price}
The feasible model prices must satisfy the individual rationality (IR) constraint \cite{mas1995microeconomic} for the buyer, i.e., bring a non-negative payoff, thereby ensuring his willingness to purchase one of the models. Accordingly, we can derive the price upper bound of each ML model $M_n$, denoted as $\overline{p}_n$ for any $n\in\mathcal{N}$, i.e.,
\begin{subnumcases}{\label{(15)}\overline p_{n}=}
&$\hspace{-9pt}U_{1}, \qquad\qquad\qquad\qquad\qquad\qquad\textrm{ }\textrm{ }\hspace{8.8pt}\textrm{ if } n=1$, \\
&$\hspace{-9pt}\frac{U_{n}+U_{1}}{2}+\sqrt{\frac{(U_{n}-U_{1})^2}{4}-\frac{(U_{n}-U_{1})C_T}{\delta^*_{n,n}-\delta^*_{n,1}}},\textrm{ if } n>1$.\label{(15b)}
\end{subnumcases}

Note that the buyer's IR constraint holds if and only if $4C_T\leq (\delta^*_{n,n}-\delta^*_{n,1})(U_{n}-U_{1})$; otherwise, the price upper bound $\overline p_{n}$ in \eqref{(15b)} does not exist.

\subsubsection{The Optimal Model Price}\label{6.2.3} 
Setting model prices to upper bound $\overline p_{n}$, as in \eqref{(15)}, maximizes the seller's payoff for each model. To optimize the seller's overall payoff, considering the buyer's strategic choices, we identify conditions for maximizing profit on any given model. This analysis determines the seller's optimal pricing scheme $\boldsymbol{p}^*$ in Stage 1 of model trading.

We first establish the condition when the sale of model $M_n$, $n\in\mathcal{N}\setminus\{1\}$, brings the seller the highest payoff, denoted as condition $\Psi_n$, where model verification must be economical and effective, as elaborated in Theorem \ref{OD}. More specifically, 
\begin{subnumcases}{\label{psi}\Psi_n:}
&$\hspace{-8pt}4C_T\leq (\delta^*_{n,n}-\delta^*_{n,1})(U_{n}-U_{1})$,\label{psia}\\
&$\hspace{-8pt}\delta^*_{n,1}(\overline p_{n}-C_{1})<\delta^*_{n,n}(\overline p_{n}-C_{n})$,\label{psib}\\
&$\hspace{-8pt} 0\leq \frac{(\overline p_{n}-C_{n})(\delta^*_{n,n}-\delta^*_{n,1})(\overline p_{n}-C_{1})}{(1-\delta^*_{n,1})(\overline p_{n}-C_{1})-(1-\delta^*_{n,n})(\overline p_{n}-C_{n})}$,\label{psic}\\
&$\hspace{-8pt}U_1-C_1 \leq \frac{(\overline p_{n}-C_{n})(\delta^*_{n,n}-\delta^*_{n,1})(\overline p_{n}-C_{1})}{(1-\delta^*_{n,1})(\overline p_{n}-C_{1})-(1-\delta^*_{n,n})(\overline p_{n}-C_{n})}$,\label{psid}\\
&$\hspace{-8pt}n=\argmax\limits_{i\in\mathcal{N}\setminus\{1\}}\frac{(\overline p_{i}-C_{i})(\delta^*_{i,i}-\delta^*_{i,1})(\overline p_{i}-C_{1})}{(1-\delta^*_{i,1})(\overline p_{i}-C_{1})-(1-\delta^*_{i,i})(\overline p_{i}-C_{i})}$.\label{psie}
\end{subnumcases}

Similarly, we can summarize the condition $\Psi_1$ in \eqref{psi_1} under which the seller's payoff of selling model $M_1$ is the highest among all models that the buyer may order, i.e., $M_i$, for all $i\in\mathcal{N}\setminus\{1\}$ that \eqref{psia} and \eqref{psib} hold.
\begin{subnumcases}{\label{psi_1}\Psi_1:}
&$\hspace{-8pt}0\leq U_1-C_1$,\\&$\hspace{-8pt}\max\limits_{i} \frac{(\overline p_{i}-C_{i})(\delta^*_{i,i}-\delta^*_{i,1})(\overline p_{i}-C_{1})}{(1-\delta^*_{i,1})(\overline p_{i}-C_{1})-(1-\delta^*_{i,i})(\overline p_{i}-C_{i})}<U_1-C_1$,
\end{subnumcases}
where $i\in\mathcal{N}\setminus\{1\}$: \eqref{psia} and \eqref{psib} hold.

Based on the discussions above, we conclude the seller's optimal pricing scheme $\boldsymbol{p}^*$ in Theorem \ref{OP}, where conditions $\Psi_n$, for all $n\in\mathcal{N}$, are mutually exclusive.

\begin{thm}
\label{OP}
The seller's optimal pricing scheme is:
\begin{itemize}
    \item if $\Psi_n$ holds: $p_n^*=\overline p_n$, $p_m^*>\overline p_m$, $\forall n\in\mathcal{N}$, $\forall m\in\mathcal{N}\setminus\{n\}$.
    \item if $\Psi_n$ does not hold, for all $n\in\mathcal{N}$: $p_n^*>\overline p_n$, $\forall n\in\mathcal{N}$.
\end{itemize}
\end{thm}





We give the detailed proof of Theorem \ref{OP} in the Appendix. Theorem \ref{OP} reveals how information asymmetry affects the optimal pricing scheme of the seller. Specifically, due to information asymmetry, the buyer suffers a payoff reduction $\Delta\pi_n$, diminishing his highest affordable model price in the trading. Such negative impact is absent only when purchasing the lowest quality model $M_1$, which is immune to model deception and thus unaffected by information asymmetry. Consequently, compared to the \textit{complete information} benchmark  (where the optimal model price equals buyer utility), information asymmetry reduces the optimal price the seller can charge for an ML model $M_n$, $n\in\mathcal{N}\setminus\{1\}$, which is lower than the buyer's corresponding model utility $U_n$, as shown in \eqref{(15b)}. 


\subsection{The Optimal Seller Payoff}
\label{S6.3}
Given the optimal pricing scheme in Theorem \ref{OP}, we can compute the resultant seller optimal payoff $\pi_{seller}^*$ below:
\begin{align}
\label{(20)}
\pi_{seller}^*=\left\{
        \begin{array}{ll}
             \hspace{-6pt}0,&\hspace{-9pt}\textrm{if $r^*=0$},\\
             \hspace{-6pt}p^*_{1}-C_{1}, &\hspace{-9pt}\textrm{if $r^*=1$},\\
             \hspace{-6pt}\frac{( p^*_{r^*}-C_{r^*})(\delta^*_{r^*,r^*}-\delta^*_{r^*,1})( p^*_{r^*}-C_{1})}{(1-\delta^*_{r^*,1})( p^*_{r^*}-C_{1})-(1-\delta^*_{r^*,r^*})( p^*_{r^*}-C_{r^*})}, &\hspace{-9pt}\textrm{if $r^*>1$},
        \end{array}
    \right.
\end{align}
where $r^*\in\{0\}\cup\mathcal{N}$. To illustrate the impact of information asymmetry on the seller, we further derive the seller's optimal payoff in the \textit{complete information} benchmark, i.e.,
\begin{align}
\label{20t}
\pi_{seller}^{benchmark}=\left\{
        \begin{array}{ll}
        0,&\textrm{if $r^*=0$},\\
        U_{r^*}-C_{r^*}, &\textrm{if $r^*\in\mathcal{N}$}.
        \end{array}
    \right.
\end{align}


By comparing the seller's optimal payoff in \eqref{(20)} and \eqref{20t}, it is surprising to see that although the seller has the opportunity of performing model deception under information asymmetry (when $r^*\in\mathcal{N}\setminus\{1\}$), her optimal payoff is less than that of the \textit{complete information} benchmark. We explain the underlying reasons for such a result as follows:
\begin{itemize}
    \item \textit{Strategic Conflict:} According to Proposition \ref{NE DV}, the distrust from information asymmetry prompts both the seller and buyer to strategically engage in model deception and model verification (with a corresponding acceptance criterion). As a result, this strategic conflict negatively impacts the seller's optimal payoff.
    \item \textit{Buyer Payoff Reduction:} Information asymmetry leads to a reduction $\Delta\pi_n$ in the buyer's payoff, which in turn lowers his maximal affordable model price. Given the individual rationality (IR) constraint of the buyer, this payoff reduction $\Delta\pi_n$ consequently diminishes the seller's optimal price and payoff in ML model trading. 
\end{itemize}

Overall, through the comparison of \textit{complete information} benchmark, information asymmetry results in a lose-lose situation for both the seller and buyer in ML model trading. To mitigate such a negative effect, we could reduce the buyer's verification cost or improve the verification accuracy, e.g., by generating synthetic test data for model verification, thereby enhancing the buyer's and seller's payoff.

So far, our analysis has focused on model trading without order information protection (OIP). The seller can strategize model delivery after observing the buyer's order, as depicted in Figure \ref{F1}. Such order information leakage enables the seller to deploy ``personalized'' (i.e., discriminatory) trading strategies, which has raised significant concerns \cite{gunderson2012amazon}. Promoted by this, we explore an alternative trading scenario in the following section, where the buyer's model order is hidden until the seller finalizes the delivery strategy. The aim is to understand how potential information protection regulations, e.g., Robinson-Patman Act \cite{hovenkamp2000robinson} and Data Governance Act \cite{Specht-Riemenschneider2025-pq}, would impact buyer and seller payoffs in model trading.

\section{Model Trading with Order Information Protection}
\label{OIP}

This section delves into model trading with order information protection (OIP). We first formally present the trading process with OIP in Section \ref{S7.1}, where the seller decides the model delivery strategy without knowing the buyer's order decision, different from that without OIP in Section \ref{SMPS}. Then, we investigate the trading Nash equilibrium through backward induction in Section \ref{7.3} and conclude the impact of OIP on the buyer and seller in Section \ref{7.4}.

\subsection{ML Model Trading Process with OIP}
\label{S7.1}
In ML model trading with OIP, we can characterize the trading process as a two-stage game, detailed as follows:
\begin{itemize}
    \item \textbf{Stage 1:} The \emph{seller} publishes a set of ML models, each with a price and specific announced quality.
    \item \textbf{Stage 2:} The \emph{seller} decides delivery strategies for each model. Meanwhile, the \emph{buyer} makes the order decision and verification strategy with an acceptance criterion.
\end{itemize}
\vspace{-8pt}\begin{figure}[ht]
    \centering
    \includegraphics[width=0.42\textwidth]{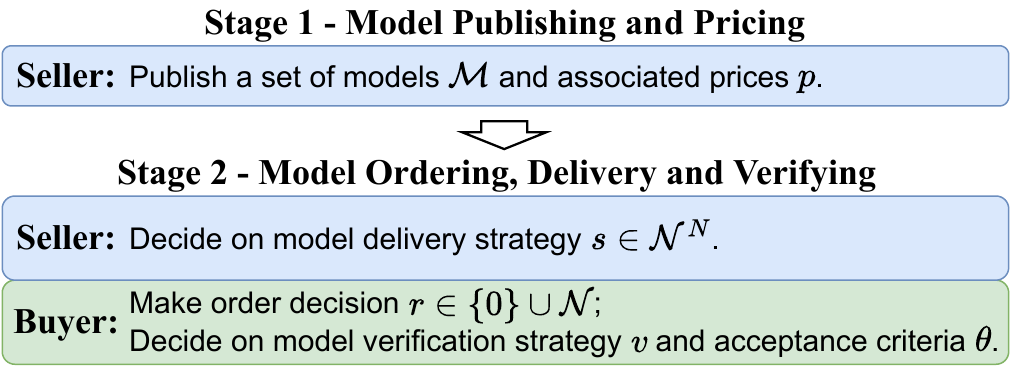}\vspace{-8pt}
    \caption{The ML Model Trading Process with Order Information Protection}
    \label{F2}
    \vspace{-3pt}
\end{figure}

In particular, for the trading interaction in Stage 2, we can further formulate it as a simultaneous game in Game \ref{Game2}.

\begin{game}[Trading Game with OIP in Stage 2]
\label{Game2}
The trading game with OIP in Stages 2 is as follows:
\begin{itemize}
    \item Players: The seller and the buyer.
    \item Strategies: The seller decides delivery strategy $\boldsymbol{s}\in\mathcal{N}^N$. The buyer makes the order decision $r\in\{0\}\cup\mathcal{N}$ and decides verification strategy $v\in\{NV,V,NA\}$, where $NA$ stands for not applicable in the case that he does not order any model. Furthermore, the buyer decides the acceptance criterion $\theta\in[0,T+1]\cup\{NA\}$, where $NA$ stands for not applicable if $v=NV$.
    \item Payoffs: The seller maximizes her expected payoff in \eqref{Revenue}. The buyer maximizes his expected payoff in \eqref{Payoff}. 
\end{itemize}
\end{game}

Note that in Stage 2, due to the unknown buyer order decision, the seller develops delivery strategy $\boldsymbol{s}\in\mathcal{N}^N$ for all models. Specifically, $\boldsymbol{s}(n)\in\mathcal{N}$  indicates the actual delivered model $M_{\boldsymbol{s}(n)}$ for the ordered model $M_n$, where $n\in\mathcal{N}$.

\subsection{Analysis of ML Model Trading with OIP}
\label{7.3}
Next, via backward induction, we investigate the equilibrium strategy in Stage 2 of model trading with OIP, as well as the optimal pricing scheme in Stage 1.

\subsubsection{Equilibrium Strategies in Stage 2} 
When dealing with Trading Game \ref{Game2} in Stage 2, we first analyze the strictly dominated strategy of the buyer, based on which we further derive the buyer's and seller's equilibrium strategy. Specifically, we divide the buyer's optimal order decision $r^*$ into the following cases to discuss the associated verification strategy.
\begin{itemize}
    \item \textit{The buyer orders nothing ($r^*=0$) in Stage 2:} When the buyer's optimal order decision is not to purchase any model, then verification and acceptance criterion decisions are not applicable, i.e., $(v^*,\theta^*)=(NA, NA)$.
    \item \textit{The buyer orders $M_r$ ($r^*\in\mathcal{N}$) in Stage 2:} If the buyer's optimal order decision is to purchase an ML model from the seller, then the corresponding buyer payoff is non-negative. Accordingly, the associated verification strategy is not to verify the model:  $(v^*,\theta^*)=(NV, NA)$, thereby avoiding extra cost and loss from imperfect verification.
\end{itemize}

From the analysis above, we can reach the buyer's dominated verification decision, as summarized in Proposition \ref{NV}.

\begin{proposition}
\label{NV}
The buyer will not verify the delivered ML model at the equilibrium of Trading Game \ref{Game2}.
\end{proposition}
We give the detailed proof of Proposition \ref{NV} in the Appendix. Following the buyer's strictly dominated strategy in Proposition \ref{NV}, the seller's best response is $\boldsymbol{s}^*(r)=1$, for each $r\in\mathcal{N}$, i.e., always deliver the lowest quality model $M_1$.

To sum up, the buyer will focus on purchasing the lowest quality model $M_1$ in Stage 2. If $U_1-p_1\geq 0$, i.e., the buyer's payoff of purchasing $M_1$ without verification is non-negative, the unique PNE of Game \ref{Game2} is $(\boldsymbol{s}^*(r^*),(r^*,v^*,\theta^*))=(1,(1,NV,NA))$. Conversely, if $U_1-p_1< 0$, the buyer will not purchase any model from the seller. We conclude the equilibrium of Trading Game \ref{Game2} in Theorem \ref{THO}.

\begin{thm}
\label{THO}
In Stage 2 of ML model trading with OIP, if $U_{1}\geq p_{1}$, the unique PNE of Game \ref{Game2} is $(\boldsymbol{s}^*(r^*),(r^*,v^*,\theta^*))=(1,(1,NV,NA))$. Otherwise, the buyer will not purchase any ML model from the seller.
\end{thm}
We give the detailed proof of Theorem \ref{THO} in the Appendix. Theorem \ref{THO} demonstrates that OIP causes ML model trading to succeed solely on the lowest quality model $M_1$. In detail, due to OIP, the buyer can prevent the loss from information asymmetry by predicting the seller's delivery strategy, without resorting to model verification. However, this approach backfires, rendering model verification an ineffective deterrent for the seller. As a result, model deception invariably occurs, leaving model $M_1$ as the only viable option for the buyer.

\subsubsection{The Optimal Pricing Scheme in Stage 1} 
Given the Nash equilibrium in Stage 2, we determine the seller's optimal pricing scheme in Stage 1, as presented in Proposition \ref{POP}. 

\begin{proposition}
\label{POP}
In Stage 1 of ML model trading with OIP, the seller's optimal pricing scheme is $p^*_1=\max\{U_1, C_1\}$ and $p^*_n= C_n$, for all $n\in\mathcal{N}\setminus\{1\}$.
\end{proposition}
We give the detailed proof of Proposition \ref{POP} in the Appendix. According to Theorem \ref{THO} and Proposition \ref{POP}, we can derive the buyer's optimal payoff $\pi_{buyer}^*$ in model trading with OIP.
\begin{align}
\label{(22)}
            \pi_{buyer}^*=\left\{
             \begin{array}{lr}
             U_{1}-p^*_{1},  &\textrm{if $U_1\geq p^*_1$},\\
             0,  &\textrm{otherwise}.
             \end{array}
        \right.
\end{align}

Similarly, with OIP, the seller's optimal payoff $\pi_{seller}^*$ is only related to the lowest quality model $M_1$, shown in \eqref{(23)}.
\begin{align}
\label{(23)}
            \pi_{seller}^*=\left\{
             \begin{array}{lr}
             p^*_{1}-C_{1},  &\textrm{if $U_1\geq p^*_1$},\\
             0,  &\textrm{otherwise}.
             \end{array}
        \right.
\end{align}

\subsection{Impact of Order Information Protection}
\label{7.4}
After analyzing the ML model trading with OIP, we explore the impact of OIP on the buyer and seller's strategy and payoff.

\subsubsection{Impact on the Buyer}
\label{7.4.1}
We first inspect how OIP affects the buyer's payoff and equilibrium trading strategy.
\begin{observation}
\label{IB1}
OIP does not increase the buyer's payoff.
\end{observation}

Observation \ref{IB1} may seem counter-intuitive, given the seller's discriminatory delivery strategy to the buyer in model trading without OIP. However, although the buyer inevitably suffers the loss from information asymmetry without OIP, the strategic model verification can mitigate such a loss and promote seller honesty. Conversely, with OIP, the buyer saves on the verification cost and evades the loss caused by imperfect verification results. Nevertheless, this makes model verification ineffective and consistently results in model deception. Consequently, OIP reduces the buyer's payoff in ML model trading.
\subsubsection{Impact on the Seller}
\label{7.4.2}
Then, we investigate the impact of OIP on the seller's strategy and payoff in model trading.

\begin{observation}
\label{IB3}
OIP does not increase the seller's payoff.
\end{observation}

For Observation \ref{IB3}, we provide the detailed explanation as follows: Without OIP, the buyer will probabilistically verify the delivered model, thereby triggering the seller's probabilistic deception strategy. In contrast, with OIP, the buyer never adopts model verification, leading the seller to perform model deception invariably. This scenario leaves the buyer interested only in the lowest quality model with OIP, preventing the seller from successfully selling higher-quality models, which could potentially yield a higher payoff.

In summary, OIP does not offer any advantages to either the buyer or the seller. Despite OIP's role in protecting the buyer from the seller's discriminatory pricing and strategy-making, it ends up being detrimental to all parties in model trading. 

Having analyzed ML model trading with and without OIP, we next turn our attention to a more diverse market where different buyers may assign varying utilities to the same model. For instance, commercial companies often derive higher utility from ML models than amateurs, as they can deploy models to enhance service quality and payoffs \cite{Ma2013-ch}. While the seller can adapt delivery strategies for different buyers to maximize payoff, the model price must remain consistent across all buyers to avoid consumer dissatisfaction \cite{Liu2021-uo}. This circumstance prompts us to design a pricing scheme that optimizes the seller's payoff in the context of model markets captured by information asymmetry and heterogeneous buyer utility.

\section{Pricing Scheme for Heterogeneous Buyers}
\label{S7}
In this section, we begin by introducing the seller's payoff optimization problem under heterogeneous buyer utility. Then, we derive the optimal pricing scheme in the ML model market without and with OIP, respectively.

\subsection{Seller Payoff Optimization Problem under Heterogeneous Buyer Utility}
Given the heterogeneity of buyer model utility, the seller aims to design a unified pricing scheme that maximizes her expected payoff. To avoid the curse of dimensionality and gain clearer insights in this initial attempt, we adopt a binary choice setting commonly used in digital products \cite{Wagner2014-co,Rao2015-vu,Liu2014-yo}, like the tiered offerings of basic and premium plans in Adobe \cite{adobe} and ChatGPT \cite{chatgpt}. Specifically, the seller will offer two types of ML models for sale: a low-quality model $M_1$ and a high-quality model $M_2$. We further assume the seller has access to the distribution (e.g., via market surveys \cite{chen2019towards,liu2020dealera,Alaei2014-cd}) of buyers' utilities for both models, i.e., utility $U_1$ of model $M_1$ and utility $U_2$ of model $M_2$ . We employ a joint probability density function $f(U_1,U_2)$ to represent heterogeneous utilities across buyers for these two models. This simplistic setup enables a deep understanding of the actual ML model market.


In each model trading, the buyer optimizes the order decision $r$ to maximize his payoff, which hinges on the model utilities $U_1$ and $U_2$, as well as the seller's pricing scheme $\boldsymbol{p}$. We thus write the buyer's optimal order decision as $r^*(\boldsymbol{p},U_1,U_2)$, and formulate the seller's payoff optimization problem under heterogeneous buyer utility as Problem \ref{P3}.

\begin{problem}\label{P3}
Seller Payoff Optimization Problem under Heterogeneous Buyer Utility.
\begin{subequations}
\begin{align}
\max&\quad \int_{U_1}^{\infty}\int_{0}^{\infty}f(U_1,U_2)\cdot\mathbb{E}[P(r^*(\boldsymbol{p},U_1,U_2))\nonumber\\&\quad-C_M(r^*(\boldsymbol{p},U_1,U_2))] \,dU_1dU_2,\\
s.t.&\quad  r^*=\argmax_{r\in\{0,1,2\}}\pi_{buyer}(r,\boldsymbol{p},U_1,U_2),\\
&\quad  0\leq p_i\leq p_j,\quad\forall i\leq j,\textrm{ }\forall i,j\in\{1,2\},\\
var.&\quad \boldsymbol{p}=\{p_1,p_2\}.
\end{align}
\end{subequations}
\end{problem}
\subsection{ML Model Market without OIP}
With the introduction of Problem \ref{P3}, we explore the model market without OIP as below. Specifically, we first conduct the theoretical analysis of Problem \ref{P3} without OIP, an inherently challenging bi-level multi-variable optimization issue (strongly NP-hard even for linear instances \cite{Hansen92}). Then, we exploit the relationship between ML model prices and the buyer’s optimal order decision, established in Section \ref{S34}-\ref{S2}, to design an efficient algorithm for solving Problem 3.

\subsubsection{Theoretical Analysis} 
\begin{figure}[ht]
    \centering
    \includegraphics[width=0.42\textwidth]{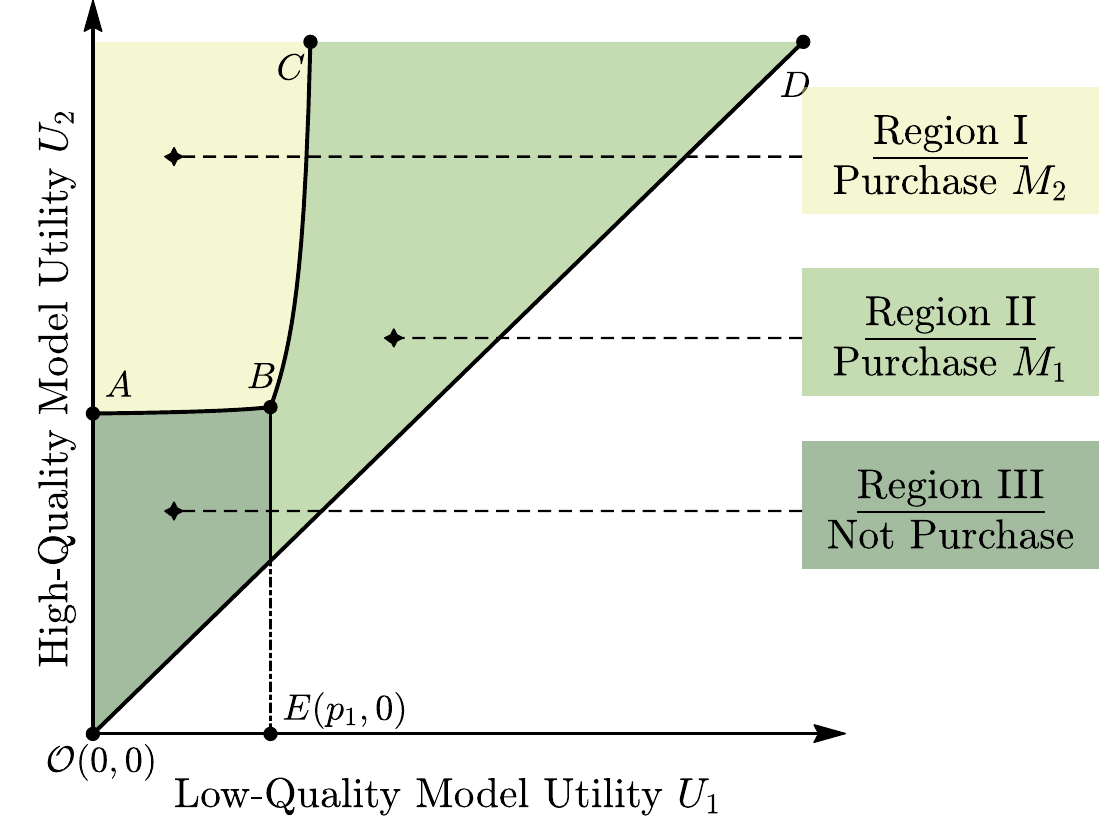}\vspace{4pt}
    \boxed{\footnotesize
\begin{array}{ll}
\text{AB:}\ U_2=(p_2^2-(\frac{C_T}{\delta^*_{2,2}-\delta^*_{2,1}}+p_2)U_1)/(p_2-U_1-\frac{C_T}{\delta^*_{2,2}-\delta^*_{2,1}}).\\
\text{BC:}\ U_2=U_1+\frac{(\delta^*_{2,2}-\delta^*_{2,1})(p_2-p_1)(U_1-p_2)}{C_T+(1-\delta^*_{2,1})(U_1-p_1)-(\delta^*_{2,2}-\delta^*_{2,1})(p_2-p_1)}.\\
\text{OD:}\ U_2=U_1.
\end{array}}\vspace{4pt}
    \caption{The Optimal Order Decision of Buyers with Heterogeneous Utility}\label{AREA}\vspace{-5pt}
\end{figure}

In model market without OIP, the seller develops discriminatory delivery strategies for heterogeneous buyers. According to Theorem \ref{OD}, Figure \ref{AREA} illustrates the distribution of buyers' optimal order decisions corresponding to model utility heterogeneity.

In Figure \ref{AREA}, line $OD$ represents buyers' minimum utility value of high-quality model $M_2$, where $U_2 = U_1$. The curves $AB$ and $BC$ denote utility thresholds for $U_1$ and $U_2$ where buyers' optimal decisions transition:
\begin{itemize}
    \item Curve $AB$: from purchasing model $M_2$ to not purchasing.
    \item Curve $BC$: from purchasing model $M_2$ to purchasing model $M_1$.
\end{itemize}

\vspace{-8pt}

The buyer's individual rationality (IR) constraint implies an economical model verification, as discussed in Section \ref{6.2}. Region \nbRoman{1} in Figure \ref{AREA} becomes feasible when verification is effective, i.e., $({\delta^*_{2,2}C_2-\delta^*_{2,1}C_1})/({\delta^*_{2,2}-\delta^*_{2,1}})<p_2$.

Our analysis reveals two distinct effects of model prices on buyers' optimal order decisions:
\begin{itemize}
    \item \textit{Monotonic Impact of $p_1$:} The low-quality model's price $p_1$ demonstrates a straightforward effect. An increase in $p_1$ shifts curve $BC$ rightward, indicating that buyers become less likely to purchase model $M_1$ and more inclined to choose the higher-quality model $M_2$.
    \item \textit{Counter-intuitive Effect of $p_2$:} The high-quality model's price exhibits a more complex influence. Interestingly, as $p_2$ increases, some buyers achieve higher payoff from model $M_2$, leading them to switch their optimal order decisions from $M_1$ to $M_2$.
\end{itemize}

Through Proposition \ref{NE DV} and the buyer's payoff reduction $\Delta\pi_2$ in \eqref{(9)}, we explain the rationale behind the impact of the high-quality model's price $p_2$. Specifically, a higher price $p_2$ mitigates the strategic conflict (i.e., model verification and deception) between the buyer and seller, decreasing the buyer's payoff reduction from information asymmetry. As a result, once this payoff reduction improvement from raising price $p_2$ exceeds the price increment, the buyer's payoff of purchasing $M_2$ increases. We derive the condition regarding model price $p_2$ and the buyer's model utilities in \eqref{(31)}, under which the buyer's payoff of purchasing $M_2$ will increase with $p_2$.
\begin{align}
\label{(31)}
    &(\delta^*_{2,2}-\delta^*_{2,1})p_2^2-2(\delta^*_{2,2}U_2-\delta^*_{2,1}U_1)p_2+(1-\delta^*_{2,1})U_1^2\nonumber\\
    &\leq(1-\delta^*_{2,2})U_2^2+(C_T-2p_2)(U_2-U_1).
\end{align}

Based on the analysis above, the variation in buyers' optimal order decisions, triggered by price $p_2$, introduces complexity to the bi-level optimization problem in Problem \ref{P3}. This variation induces non-convexity, making it difficult to derive a closed-form solution. Therefore, in the following, our focus shifts towards developing an efficient algorithm for Problem \ref{P3}.

\subsubsection{Algorithm Design} 
When tackling such a bi-level non-convex optimization problem, inspired by finite element methods \cite{Bangerth2003-hy,Babuska2001-lv,Szabo2021-ml}, we decompose Problem \ref{P3} into more tractable sub-problems, thus designing Algorithm \ref{Algo1}. In particular, we integrate the monotonic correlation between model price $p_1$ and optimal order decisions within Algorithm \ref{Algo1}. This integration enables us to enhance computational efficiency by reducing the search space to a single dimension compared to an exhaustive search. The detailed steps of Algorithm \ref{Algo1} are as follows:

\emph{Workflow of Algorithm \ref{Algo1}:} Without loss of generality, we first define $U$ as the utility upper bound of buyers' model utility distribution. Then, for each feasible price $p_2$ (Line $2$) that ensures non-negative payoffs for buyers purchasing model $M_2$, we identify two threshold prices for $p_1$:
\begin{subnumcases}{\label{etath}}
    \eta_e(p_2)&$\hspace{-8pt}=\frac{(\delta^*_{2,2}-\delta^*_{2,1})(p_2-C_2)p_2+(1-\delta^*_{2,1})(C_2-C_1)C_1}{(1-\delta^*_{2,1})(p_2-C_1)-(1-\delta^*_{2,2})(p_2-C_2)}$,\\
    \eta_h(p_2)&$\hspace{-8pt}=\frac{p_2^2- (p_2-{C_T}/({\delta^*_{2,2}-\delta^*_{2,1})}) U}{p_2- U+{C_T}/({\delta^*_{2,2}-\delta^*_{2,1}})}$.
\end{subnumcases}
Specifically, we derive the threshold prices $\eta_e(p_2)$ and $\eta_h(p_2)$ in \eqref{etath} from the perspective of the seller and buyer (Line $3$), respectively. Given the high-quality model's price $p_2$, $\eta_e(p_2)$ represents the threshold price of $p_1$ at which the seller achieves equivalent payoffs from selling either model. In contrast, $\eta_h(p_2)$ is the threshold price of $p_1$ that leads the optimal order decision to fall into Region \nbRoman{1} in Figure \ref{AREA}, for any buyer who obtains a non-negative payoff from purchasing $M_2$.

According to threshold prices in \eqref{etath}, we determine the seller's optimal price $ p_1$ corresponding to different price $p_2$ (Lines $4$-$10$). We divide the investigation into the following two scenarios based on buyer payoff in model trading.

\emph{Scenario 1: buyers can obtain a non-negative payoff from both models.} To optimize price $p_1$, we analyze the response of the seller's expected payoff to variations in price $p_1$ relative to threshold $\eta_e$, denoted as the difference $\Delta{p_1}=p_1-\eta_e$. By comparing threshold prices $\eta_e(p_2)$ and $\eta_h(p_2)$, we quantify how expected seller payoff varies with $\Delta{p_1}$, as shown in \eqref{FF}.
\begin{subnumcases}{\label{FF}}
    \hspace{-3pt}F_{1}\hspace{-2pt}=\hspace{-8pt}&$\Delta{p_1}(\int_{\eta_e+\Delta{p_1}}^\eta\int_{U_1}^BfdU_2dU_1+\int_{\eta}^{U}\int_{U_1}^{U}fdU_2dU_1)$\notag\\
    &$-(\eta_e-C_1)\int_{\eta_e}^{\eta_e+\Delta{p_1}}\int_{U_1}^{A}fdU_2dU_1$,\label{FF1}\\
    \hspace{-3pt}F_{2}\hspace{-2pt}=\hspace{-8pt}&$\Delta{p_1}(\int_{\eta}^{\eta_h}\int_{A}^UfdU_2dU_1+\int_{\eta_e}^U\int_{U_1}^{U} fdU_2dU_1$\notag\\
    &$+\int_{\eta_e+\Delta{p_1}}^\eta\int_{A}^BfdU_2dU_1)+(\eta_e+\Delta{p_1}-C_1)$\notag\\
    &$(\int_{\eta_e+\Delta{p_1}}^{\eta_h}\int_{U_1}^{A}fdU_2dU_1+\int_{\eta_h}^{\eta_e}\int_{U_1}^UfdU_2dU_1)$,\label{FF2}
\end{subnumcases}
where $A(U_1)$ and $B(U_1,\Delta{p_1})$ respectively denote the curve functions (with $U_2$ as the dependent variable) $AB$ and $BC$ in Figure \ref{AREA}. \begin{algorithm}
    \label{Algo1}
    \caption{Seller Payoff Optimization under Heterogeneous Buyer Utility}
    \KwIn{model costs $C_1$ and $C_2$, model utility upper bound $U$, verification cost $C_T$, acceptance probability $\delta^*_{2,1}$ and $\delta^*_{2,2}$, and step size $\alpha$.}
    \KwOut{the optimal pricing scheme $\boldsymbol{p}^*=\{p_1^*,p_2^*\}$ and the resultant expected payoff $\pi_{seller}^*$.}
    $\pi_{seller}^*=0$\;
    \For{$p_2=\frac{\delta^*_{2,2}C_2-\delta^*_{2,1}C_1}{\delta^*_{2,2}-\delta^*_{2,1}}:\alpha:\frac{1}{2}( U+\sqrt{ U^2-\frac{4 U C_T}{\delta^*_{2,2}-\delta^*_{2,1}}})$}{Calculate $\eta_e(p_2)$ and $\eta_h(p_2)$ in \eqref{etath}\;
\eIf{$\eta_e(p_2)\leq\eta_h(p_2)$}{$F(\Delta{p_1})\leftarrow F_1(\Delta{p_1})$ in \eqref{FF1}\;
}{$F(\Delta{p_1})\leftarrow F_2(\Delta{p_1})$ in \eqref{FF2}\;}
$\Delta{p_1}=\{x\in[C_1-\eta_e,\eta_h-\eta_e]:F'(x)=0\}$\;
$\hat{p_1}=\{x\in(\eta_h,U):G'(x)=0\}$\;
$p_1=\arg\max_{p_1\in\{\eta_e+\Delta{p_1},\hat{p_1}\}}\pi_{seller}(p_1,p_2)$\;
\If{$\pi_{seller}(p_1,p_2)>\pi_{seller}^*$}{$\boldsymbol{p^*}=(p_1,p_2)$\;
$\pi_{seller}^*=\pi_{seller}(p_1,p_2)$\;}
}

Return $\boldsymbol{p^*}$ and $\pi^*_{seller}$\;
\end{algorithm}Moreover, $\eta$ corresponds to the point $C$ in Figure \ref{AREA}, which is related to the utility upper bound $U$.

\emph{Scenario 2: buyers can only obtain a non-negative payoff from one of the models.} 
In this case, the seller optimizes price $p_1$ within the range $(\eta_h,U)$, regardless of the impact on buyers purchasing model $M_2$. We thus focus on the seller's expected payoff from selling model $M_1$, as derived in \eqref{GF}.
\begin{align}
    G({p_1})=({p_1}-C_1)\int_{{p_1}}^U \int_{U_1}^Uf(U_1,U_2)dU_2dU_1.\label{GF}
\end{align}

After the preceding analysis, we determine the potential set of optimal pricing schemes by calculating the critical points (Lines $8$-$9$) of the functions in \eqref{FF} and \eqref{GF}. With iteratively updating the pricing scheme that maximizes expected seller payoff (Lines $10$-$13$), Algorithm \ref{Algo1} outputs the optimal pricing scheme and resultant expected payoff of the seller.
\subsection{ML Model Market with OIP}
Next, we delve into the ML model market with OIP under heterogeneous buyer utility. From Theorem \ref{THO}, due to OIP, the low-quality model $M_1$ is the only option buyers consider for purchasing in model trading. Consequently, the seller aims to determine the optimal price $p_1^*$ that maximizes her expected payoff $\pi_{seller}(p_1)$ in \eqref{SP_oip}.
\begin{align}\label{SP_oip}
    \pi_{seller}(p_1)=(p_1-C_1)\int_{p_1}^\infty\int_{0}^{\infty} f(U_1,U_2)dU_2dU_1.
\end{align}

The method of deriving $p_1^*$ is similar to that of optimizing $G(p_1)$ in \eqref{GF}, where the closed-form solution depends on the specific distribution of buyers' model utility. For example, when the probability density function of model utility $f(U_1,U_2)$ follows a uniform distribution within interval $[0,U]$, the optimal price is $p_1^*=(2C_1+U)/3$. Therefore, we conclude that the seller's optimal pricing scheme is $p_1^*=\argmax_{p_1}(p_1-C_1)\int_{p_1}^\infty\int_{0}^{\infty} f(U_1,U_2)dU_2dU_1$ and $p_2^*= C_2$. 

\section{Numerical Results}
\label{EXP}
This section provides numerical results to evaluate our equilibrium analysis in ML model trading. In particular, we demonstrate how different system parameters and ML model trading scenarios affect the payoff of the seller and buyer.

\subsection{Experimental Setup}

\subsubsection{The Seller}
To better characterize the actual ML model market, we use the widely adopted Resnet-18 model \cite{He2016-dd} and CIFAR-10 \cite{krizhevsky2009learning} data set in our study, which comprises $60,000$ color images of objects. For simplicity, we focus on a set of $N=5$ models for sale, each with distinct qualities. The training data size varies across these models, with a large data set yielding better model quality, as summarized in Table \ref{CF10}. We further denote $Q_n$ as the model quality of $M_n$, for any $n\in\mathcal{N}$, which correlates with the test accuracy, e.g., $Q_5=95.29\%$. Moreover, we designate the model cost as the $1\%$ of the training data size, e.g., $C_5=500$.
\vspace{-8pt}\begin{table}[ht]
\caption{ML models with different training data sizes}\vspace{-5pt}
\label{CF10}
    \centering
    \begin{tabular}{|c|c|c|c|c|c|}
    \hline
	\textbf{ML Model} &$M_1$ &$M_2$&$M_3$&$M_4$&$M_5$ \\
    \hline
	\textbf{Training Data Size} & $10000$ & $20000$  & $30000$ &$40000$ &$50000$ \\ \hline
	\textbf{Accuracy $Q_n$ ($\%$)} & $85.90$ & $92.05$  & $93.84$& $94.97$& $95.29$\\ \hline
	\textbf{Cost $C_n$} & $100$ & $200$  & $300$&$400$&$500$\\ \hline
	\textbf{Concave Utility $U_n$}& $120$ & $244$  & $261$&$271$&$273$\\ \hline
    \textbf{Convex Utility $U_n$} & $120$ & $309$  & $435$&$531$&$560$\\ \hline
    \end{tabular}
    \vspace{-5pt}
\end{table}
\subsubsection{The Buyer} We explore two different types of buyer utility functions regarding model quality $Q_n$: concave and convex. We initialize the utility of the lowest quality model $M_1$ to $120$, and define the buyer's utility functions as follows:
\begin{itemize}
    \item Concave function: $U_n=500(Q_n-Q_1)^{0.5}+120$,
    \item Convex function: $U_n=50000(Q_n-Q_1)^2+120$.
\end{itemize}
\subsection{Buyer Payoff under Different Scenarios}
We first evaluate the buyer's optimal payoff under different scenarios, where we adopt a proportional pricing scheme that $p_n=1.1C_n$, $n\in\mathcal{N}$, for ease of exposition.

\subsubsection{Impact of Acceptance Criterion and Test Data Size} 
\begin{figure}[ht]
    \centering
    \subfigure[Buyer Payoff with Concave Utility]{\label{Bt_a}\includegraphics[width=0.24\textwidth]{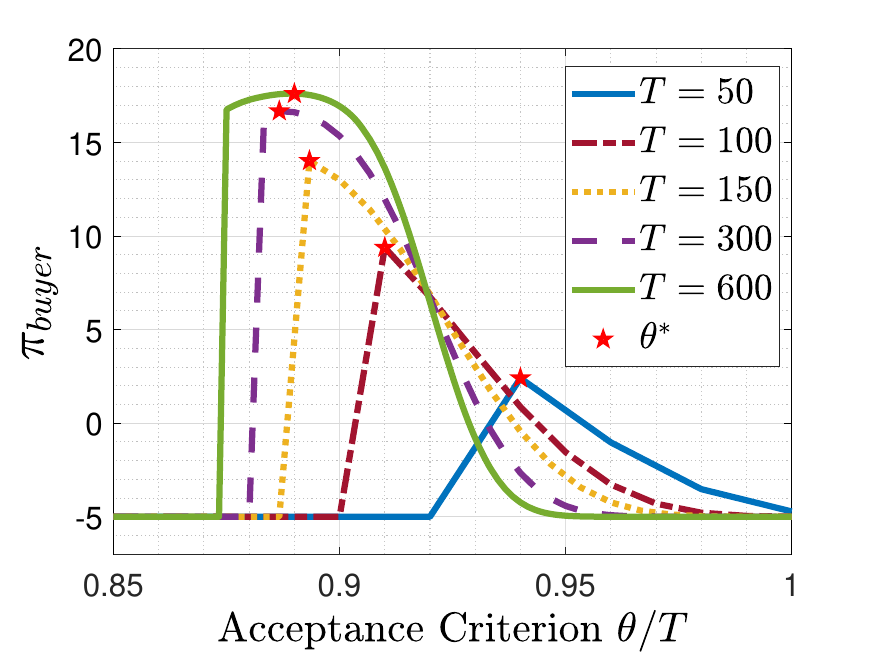}}
    \subfigure[Buyer Payoff with Convex Utility]{\label{Bt_v}\includegraphics[width=0.24\textwidth]{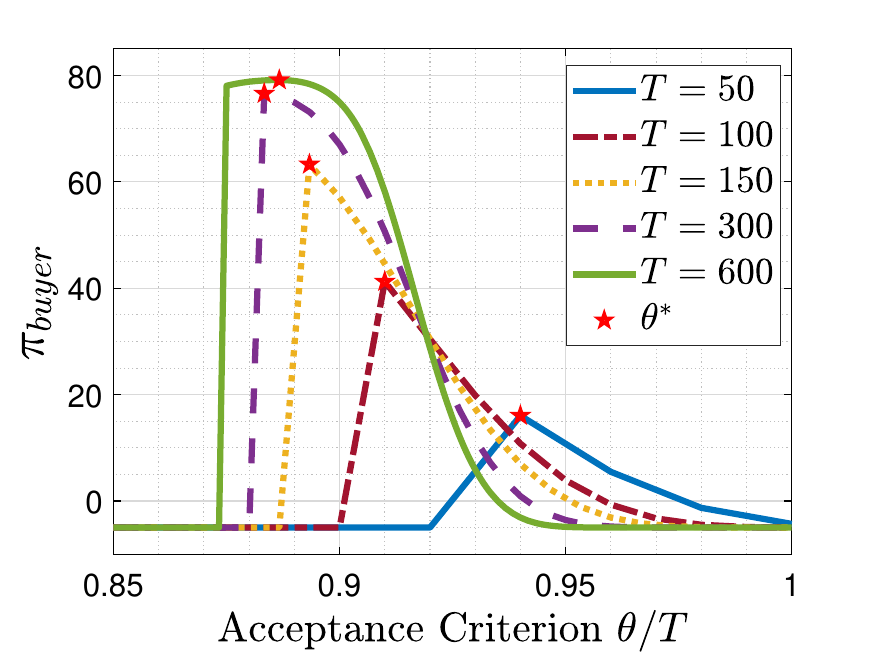}}
    \vspace{-12pt}
    \caption{Buyer Payoff under Different Acceptance Criterion}
    \vspace{-12pt}
    \label{BP1}
\end{figure}

To investigate how acceptance criterion $\theta$ and test data size $T$ affect the buyer's payoff in model trading without OIP, we take verification cost $C_T=5$ and order decision $r=2$ in Figure \ref{BP1} as an illustrative example. Specifically, we first study the impact of acceptance criterion $\theta$ on buyer payoff. In both subfigures, the buyer achieves the maximal payoff under the acceptance criterion $\theta^*$, which aligns with Proposition \ref{NE DV}.

Then, we illustrate the impact of test data size $T$ on the buyer. From Figure \ref{BP1}, the buyer's optimal payoff increases with test data size $T$, as summarized in Observation \ref{bpint}. The reason is that a larger test data size $T$ improves the buyer's accuracy in determining model quality through verification.
\begin{observation}\label{bpint}
    The buyer's optimal payoff increases with test data size in model trading without OIP.
\end{observation}
Furthermore, with a larger test data size $T$, the buyer lowers the optimal acceptance criterion $\theta^*$ to maximize his payoff, as presented in Observation \ref{tila}. In other words, as $T$ increases, the buyer accepts greater quality fluctuations in model verification. The underlying rationale is also that the larger test data size decreases the probability of overestimating or underestimating the delivered model quality due to imperfect verification.
\begin{observation}\label{tila}
    As test data size increases, the buyer lowers the acceptance criterion $\theta^*$ in model verification without OIP.
\end{observation}
\subsubsection{Impact of Verification Cost-effectiveness} 
\begin{figure}[ht]
\vspace{-12pt}
    \centering
    \subfigure[Buyer Payoff with Concave utility]{\label{Bp_a}\includegraphics[width=0.24\textwidth]{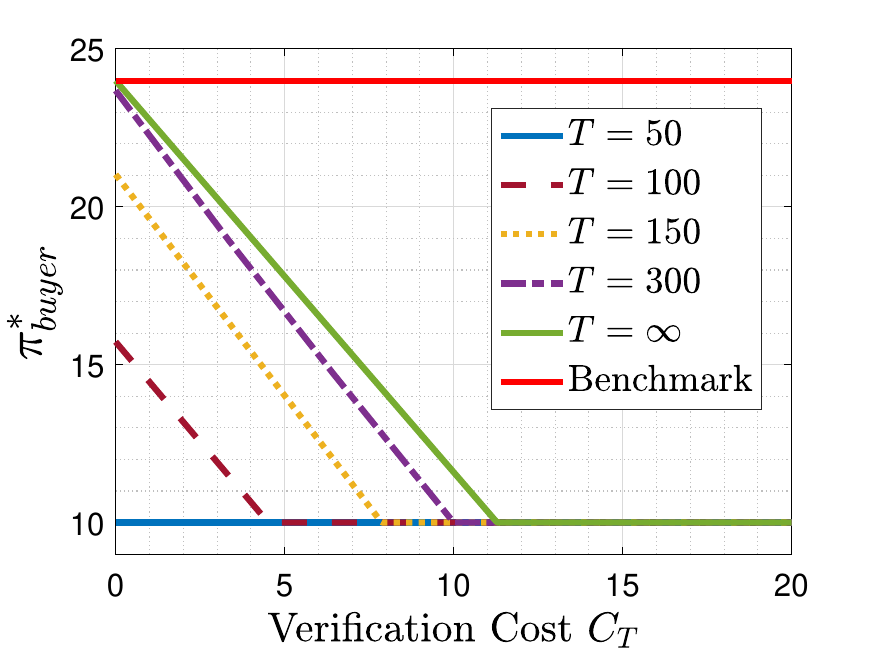}}
    \subfigure[Buyer Payoff with Convex utility]{\label{Bp_v}\includegraphics[width=0.24\textwidth]{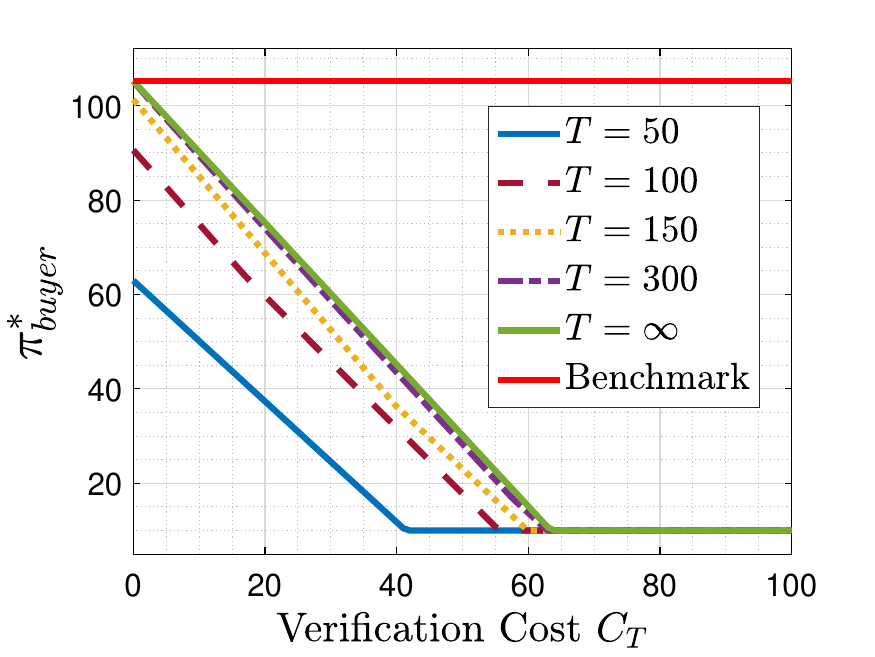}}
    \vspace{-12pt}
    \caption{The Optimal Buyer Payoff under Different Verification Cost}
    \vspace{-4pt}
    \label{BP2}
\end{figure}

In Figure \ref{BP2}, we explore the impact of verification cost-effectiveness on the buyer's optimal payoff and compare it with the \textit{complete information} benchmark. As shown in both subfigures, the buyer's optimal payoff decreases with the verification cost $C_T$, which is consistent with our analysis of model trading without OIP in Section \ref{S2}. We summarize this result in Observation \ref{opv}.
\begin{observation}\label{opv}
    The buyer's optimal payoff decreases with the verification cost in model trading without OIP.
\end{observation}
Additionally, from Figure \ref{BP2}, a larger test data size enables the buyer to bear a higher $C_T$ for purchasing high-quality models, as elaborated in Observation \ref{tibh}. The threshold for $C_T$ (where the buyer’s payoff reaches $10$), above which the buyer opts for the lowest quality model, rises with test data size.
\begin{observation}\label{tibh}
    As test data size increases, the buyer can afford higher verification costs in model trading without OIP.
\end{observation}

\subsubsection{Impact of Order Information Protection} 
\begin{figure}[ht]
    \centering
    \subfigure[Buyer Payoff with Concave Utility]{\label{Bp_ao}\includegraphics[width=0.24\textwidth]{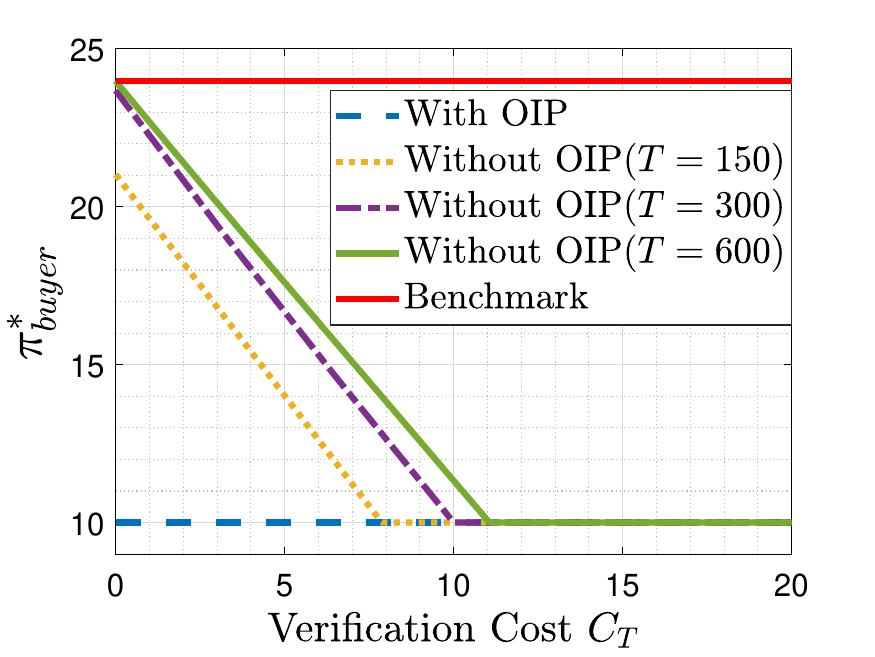}}
    \subfigure[Buyer Payoff with Convex Utility]{\label{Bp_vo}\includegraphics[width=0.24\textwidth]{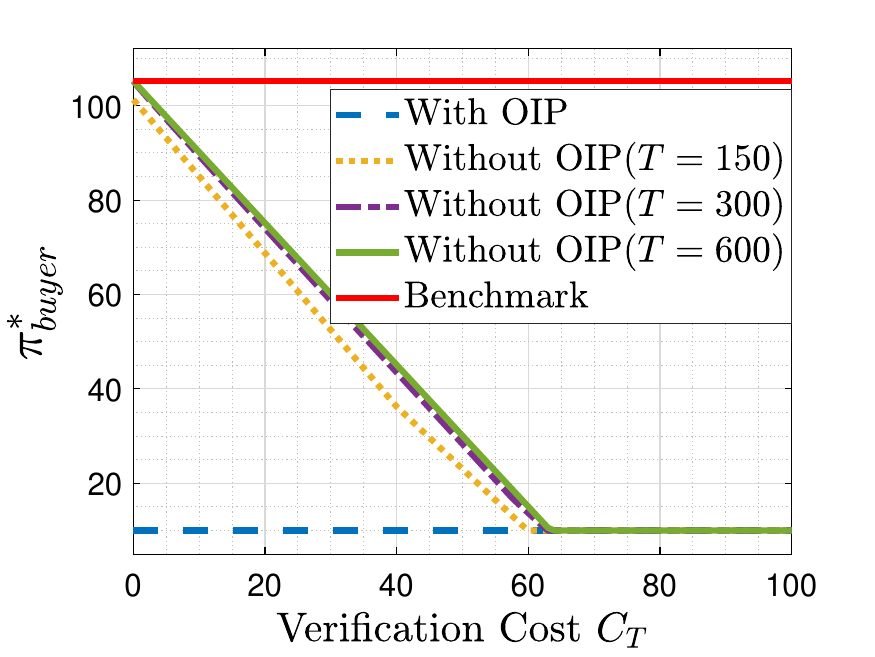}}
    \vspace{-12pt}
    \caption{The Optimal Buyer Payoff With and Without OIP}
    \vspace{-12pt}
    \label{BP3}
\end{figure}

Figure \ref{BP3} illustrates the impact of order information protection (OIP) on the buyer's optimal payoff. In both subfigures, OIP results in the minimum constant buyer payoff. In contrast, without OIP, a more cost-effective model verification (i.e., lower verification costs and more test data) significantly enhances the buyer's optimal payoff. In particular, when verification cost $C_T=0$ without OIP, merely a few hundred test data, e.g., $T=300$, are sufficient to achieve the optimal buyer payoff nearly equivalent to that under the \textit{complete information} benchmark. We summarize the above observations in Observation \ref{boipwoip}.

\begin{observation}\label{boipwoip}
    OIP leads to the minimum constant buyer optimal payoff. Conversely, without OIP, the buyer's optimal payoff increases with the cost-effectiveness of verification, eventually aligning with the complete information benchmark.
\end{observation}
\subsection{Seller Payoff under Different Scenarios}
\label{ExpC}
Next, we explore the seller's optimal payoff under different scenarios with our proposed pricing scheme in Theorem \ref{OP} and summarize the experimental results in Observation \ref{spds}.

\subsubsection{Impact of Verification Cost-effectiveness}
\begin{figure}[ht]
\vspace{-12pt}
    \centering
    \subfigure[Seller Payoff with Concave Utility]{\label{Sp_a}\includegraphics[width=0.24\textwidth]{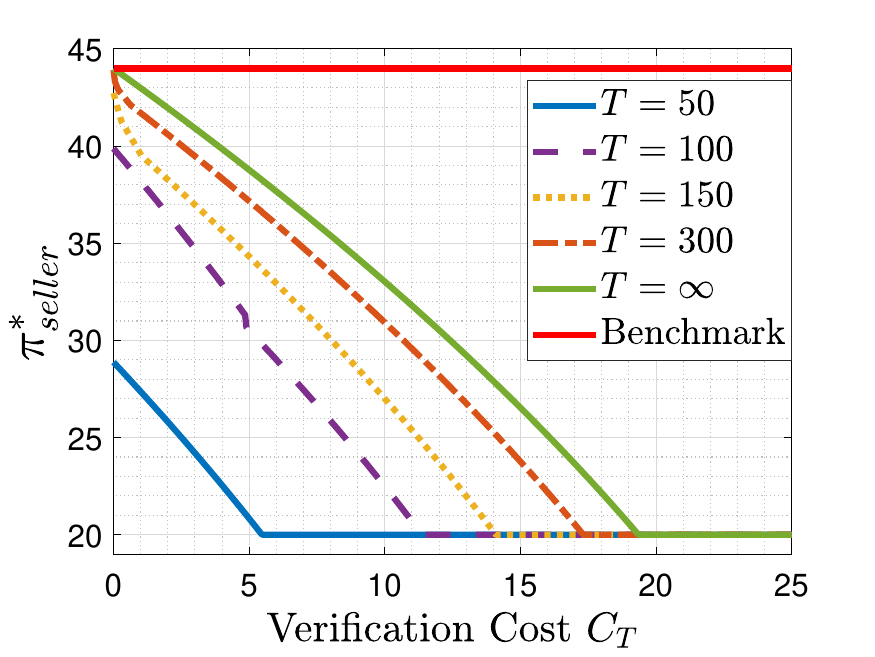}}
    \subfigure[Seller Payoff with Convex Utility]{\label{Sp_v}\includegraphics[width=0.24\textwidth]{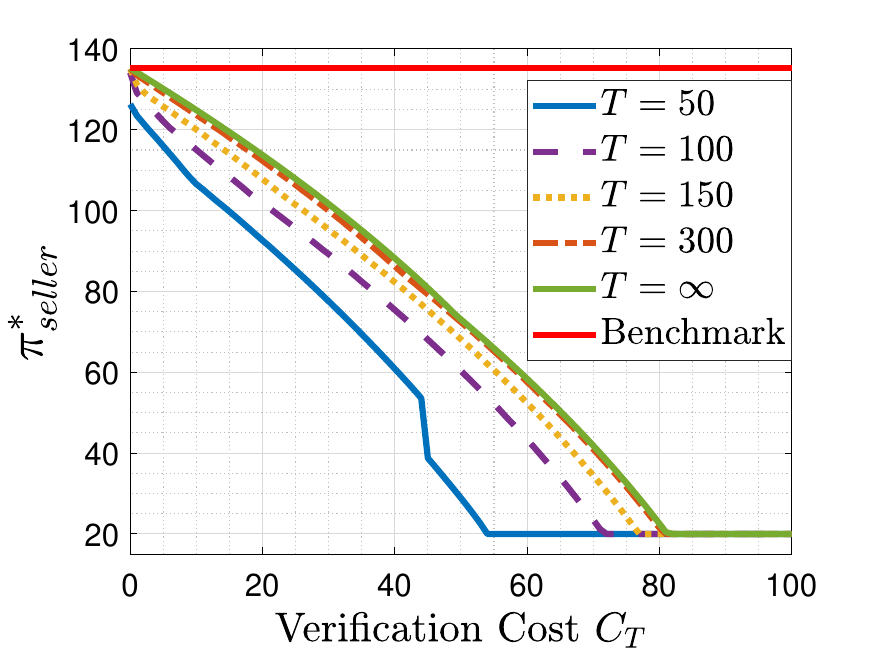}}
    \vspace{-12pt}
    \caption{The Optimal Seller Payoff under Different Verification Cost}
    \vspace{-6pt}
    \label{SP1}
\end{figure}

In Figure \ref{SP1}, the seller's payoff under our proposed optimal pricing scheme increases with test data size and decreases with verification cost, which is consistent with our discussions in Section \ref{S6.3}.
\subsubsection{Impact of Order Information Protection} 
\begin{figure}[ht]
\vspace{-12pt}
    \centering
    \subfigure[Seller Payoff with Concave Utility]{\label{Sp_ao}\includegraphics[width=0.24\textwidth]{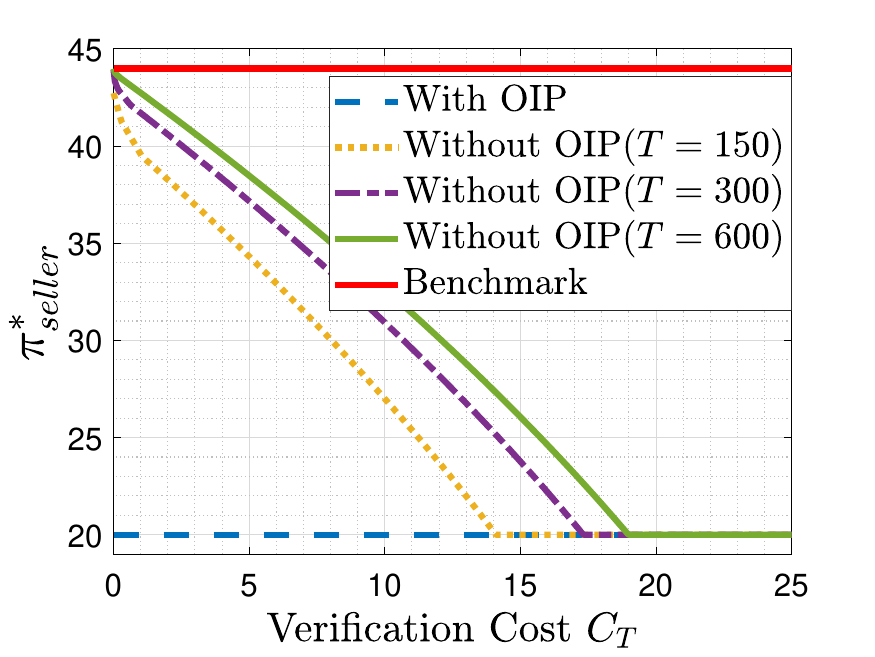}}
    \subfigure[Seller Payoff with Convex Utility]{\label{Sp_vo}\includegraphics[width=0.24\textwidth]{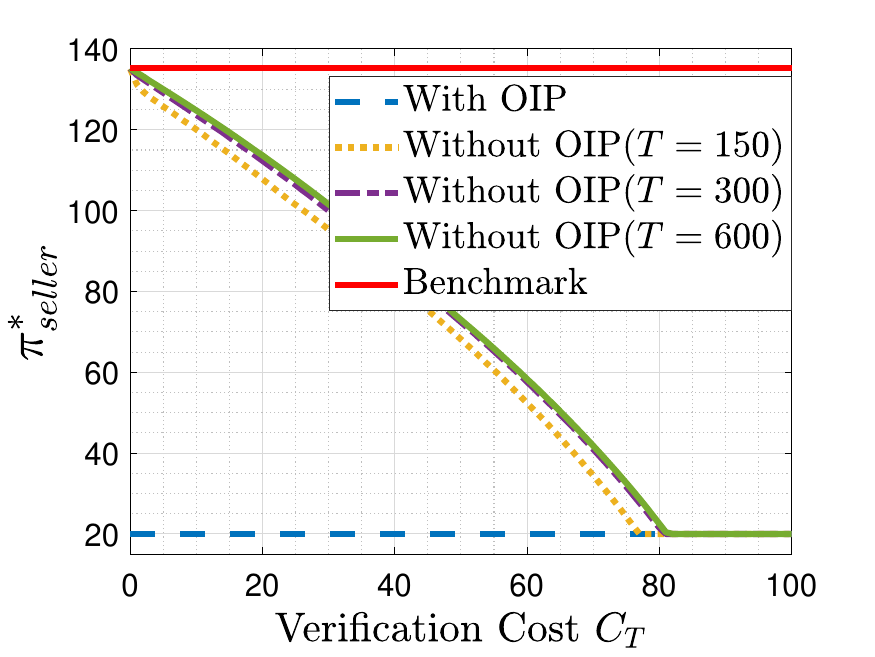}}
    \vspace{-12pt}
    \caption{The Optimal Seller Payoff With and Without OIP}
    \vspace{-8pt}
    \label{SP2}
\end{figure}

In Figure \ref{SP2}, OIP reduces the seller's optimal payoff, except when the buyer purchases the lowest quality model (unaffected by information asymmetry), where the resultant optimal payoff is always $20$.
\begin{observation}\label{spds}
    OIP leads to the minimum constant seller optimal payoff. Conversely, without OIP, the seller's optimal payoff increases with the cost-effectiveness of verification, eventually aligning with the complete information benchmark.
\end{observation}
\subsection{Expected Seller Payoff under Heterogeneous Buyer Utility}
We further inspect the seller's expected payoff under heterogeneous buyer utility to evaluate our proposed pricing scheme in Section \ref{S7}. Specifically, we adopt a uniform distribution of buyers' model utility within the interval $[0, U]$. The insights for other distribution types of model utility are the same.
\subsubsection{Impact of Order Information Protection and Verification Cost-effectiveness}
Figure \ref{Sp_hc} illustrates the impact of OIP and verification cost-effectiveness on the seller's expected payoff under heterogeneous buyer utility, where ${U}=300$. Similar to Observation \ref{spds}, a more cost-effective verification also increases the seller's expected payoff under heterogeneous buyer utility without OIP. Meanwhile, OIP leads to the minimum constant seller's expected payoff.

\subsubsection{Impact of Buyer Utility Distribution}
\begin{figure}[ht]
\vspace{-8pt}
    \centering
    \subfigure[Seller Payoff with Different Verification Cost]{\label{Sp_hc}\includegraphics[width=0.24\textwidth]{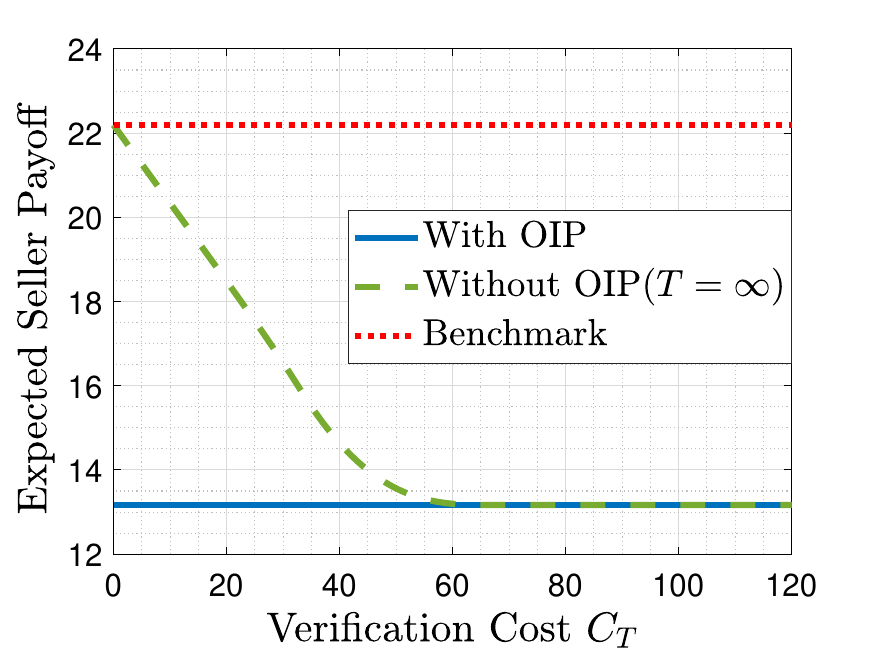}}
    \subfigure[Seller Payoff with Different Upper Bound of Model Utility Distribution]{\label{Sp_hu}\includegraphics[width=0.24\textwidth]{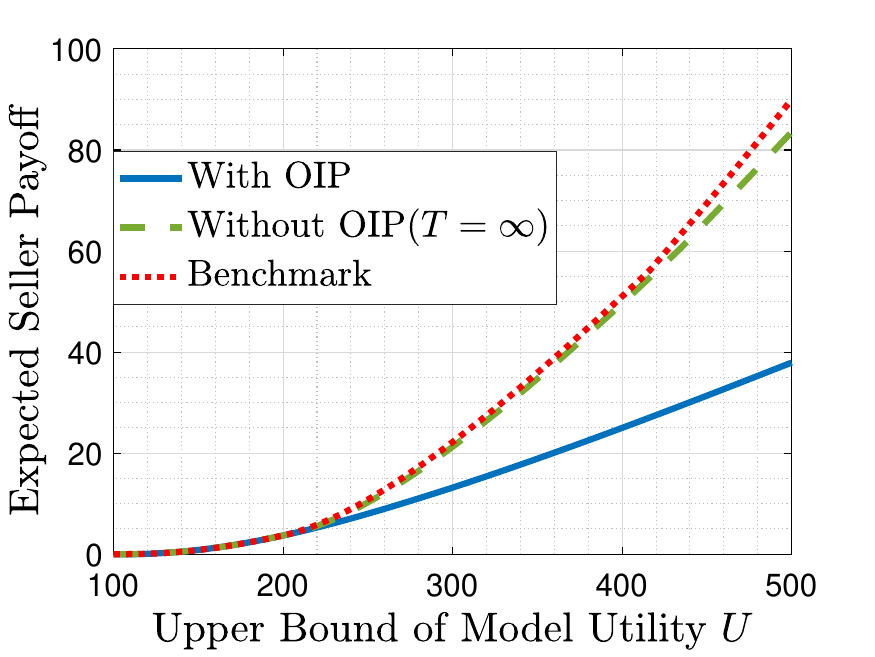}}
    \vspace{-12pt}
    \caption{The Optimal Seller Payoff under Uniform Buyer Utility Distribution}
    \vspace{-6pt}
    \label{SP3}
\end{figure}

Figure \ref{Sp_hu} investigates the impact of buyer utility distribution on the seller's expected payoff with a fixed verification cost $C_T=5$. As the upper bound $U$ of utility distribution escalates, indicating more buyers with higher model utility in the market, the seller's expected payoff in all scenarios (i.e., with and without OIP, and complete information benchmark) also increases. Meanwhile, the disparity among the three curves in Figure \ref{Sp_hu} widens with $U$. Notably, our proposed pricing scheme achieves the seller's expected payoff without OIP, nearly equating to the complete information benchmark and much higher than that with OIP. Only when the upper bound of utility distribution is insufficiently high (i.e., $U\leq 200$) to enable the seller to obtain a positive payoff from selling high-quality models, the seller's expected payoff of the three curves aligns. This is because, in such cases, the seller maximizes her payoff by only selling the lowest quality model, unaffected by information asymmetry. 

\section{Conclusion}
\label{Con}
Our research is the first to systematically study ML model trading under information asymmetry. By using dominated strategy elimination and learning curves' power-law feature, we uncover that accurate, cost-effective verification benefits both buyers and sellers by reducing information asymmetry. We also investigate how order information protection (OIP) unexpectedly lowers both buyer and seller payoffs by removing discriminatory trading strategies.

To optimize the seller's expected payoff with and without OIP, we introduce a pricing strategy tailored to diverse buyer utilities, approximating the complete information benchmark. Future work will expand this pricing strategy to multiple models and refine our model to account for complex cost-performance correlations, moving beyond the power-law feature of learning curves. These advancements will make our findings more practical and widely applicable, despite complicating the equilibrium analysis and pricing strategy in the presence of information asymmetry.

\section*{Acknowledgment}
This work is supported by National Natural Science Foundation of China (Project 62271434, 12371519, 61771013), Fundamental Research Funds for the Central Universities of China, Shenzhen Key Lab of Crowd Intelligence Empowered Low-Carbon Energy Network (No. ZDSYS20220606-100601002), Shenzhen Stability Science Program 2023, Shenzhen Institute of Artificial Intelligence and Robotics for Society, and Longgang District Shenzhen's "Ten Action Plan" for Supporting Innovation Projects (No. LGKCSDPT2024002).

%



\ifCLASSOPTIONcaptionsoff
  \newpage
\fi



%
\bibliographystyle{IEEEtran}
\bibliography{ref}

%

\begin{IEEEbiography}[{\includegraphics[width=1in,height=1.25in,clip,keepaspectratio]{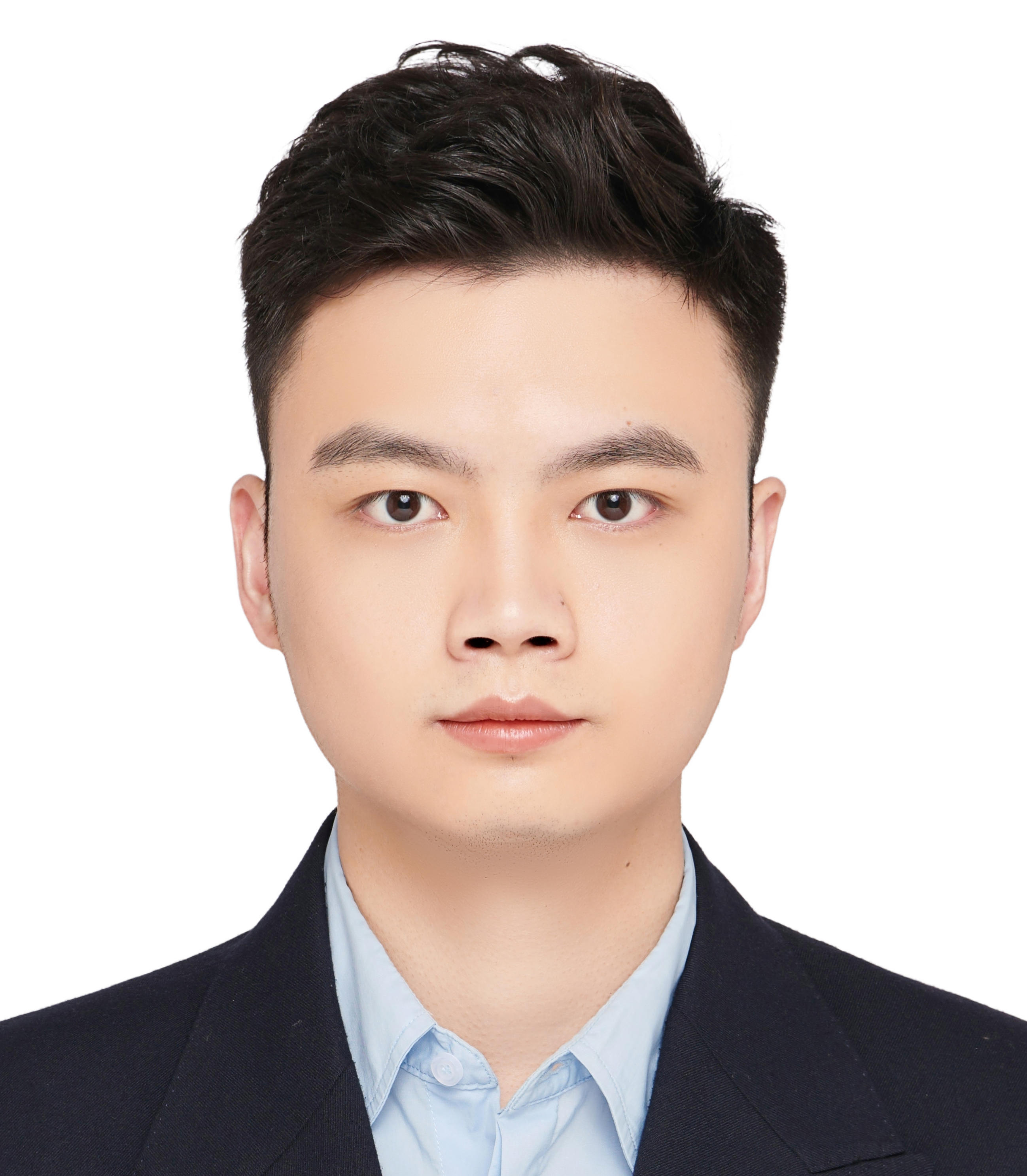}}]{Xiang Li} (Student Member, IEEE) received the B.S. degree from the University of Electronic Science and Technology of China, in 2020. He is currently pursuing the Ph.D. degree with the School of Science and Engineering, The Chinese University of Hong Kong, Shenzhen. His research interests include
the field of network economics and optimization, with current emphasis on pricing and mechanism design in the context of artificial intelligence.
\end{IEEEbiography}

\begin{IEEEbiography}[{\includegraphics[width=1in,height=1.25in,clip,keepaspectratio]{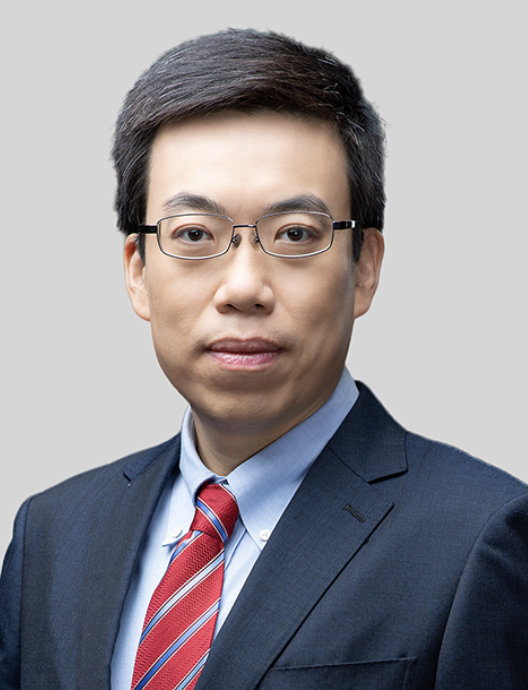}}]{Jianwei Huang} (Fellow, IEEE) is a Presidential Chair Professor and Associate Vice President at the Chinese University of Hong Kong, Shenzhen. He is also the Associate Director of the Shenzhen Institute of Artificial Intelligence and Robotics for Society. He has served as the Editor-in-Chief of IEEE Transactions on Network Science and Engineering and the Associate Editor-in-Chief of the IEEE Open Journal of the Communications Society. He is an IEEE Fellow, an IEEE ComSoc Distinguished Lecturer, a Clarivate Web of Science Highly Cited Researcher, and an Elsevier Most Cited Chinese Researcher.
\end{IEEEbiography}

\begin{IEEEbiography}[{\includegraphics[width=1in,height=1.25in,clip,keepaspectratio]{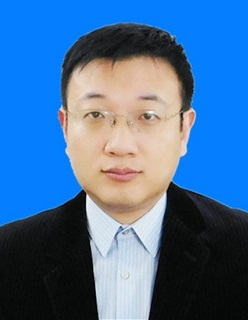}}]{Kai Yang} (Senior Member, IEEE) received the Ph.D. degree from Columbia University, USA. He has held various prestigious positions, including Technical Staff Member at Bell Laboratories, Alcatel-Lucent, USA. He has also served as an adjunct faculty member at Columbia University. Currently, he is a Distinguished Professor at Tongji University, Shanghai, China. He holds over 20 patents and has been extensively published in Top AI conferences and leading IEEE journals and conferences. His current research interests encompass AI+Networking, AI for Time Series, Nested Optimization, Wireless Sensing and Communications. He has also served as a Technical Program Committee (TPC) member for numerous AI and IEEE conferences. He has received several prestigious awards, including the Eliahu Jury Award from Columbia University, the Bell Laboratories Teamwork Award, and the Best Paper Honorable Mention Award for time series workshop of IJCAI 2023. From 2012 to 2014, he was the Vice-Chair of the IEEE ComSoc Multimedia Communications Technical Committee. Additionally, he has held leadership roles such as Demo/Poster Co-Chair for IEEE INFOCOM, Symposium Co-Chair for IEEE GLOBECOM, and Workshop Co-Chair for IEEE ICME.
\end{IEEEbiography}

\begin{IEEEbiography}[{\includegraphics[width=0.9in,height=1.25in,clip,keepaspectratio]{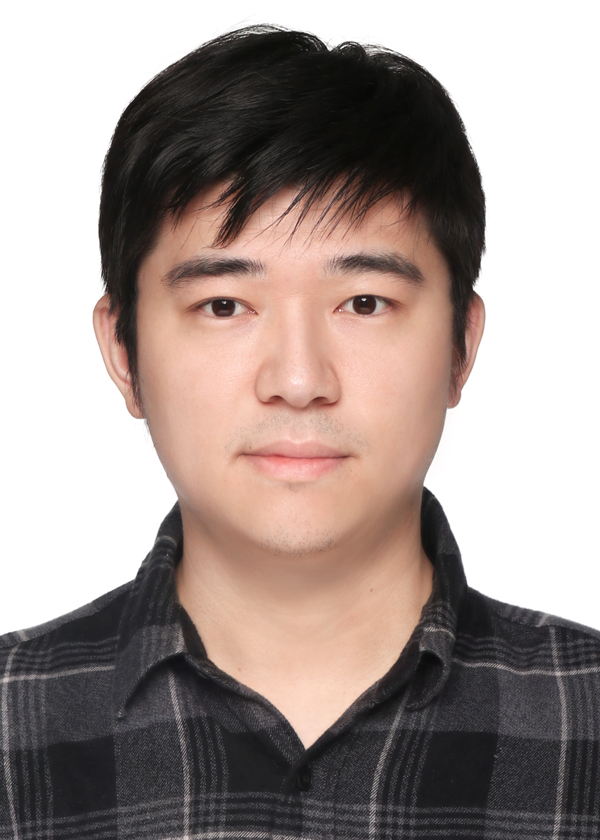}}]{Chenyou Fan} (Member, IEEE)
received the B.S. degree in computer science from the Nanjing University, China, in 2011, and the M.S. and Ph.D. degrees from Indiana University, USA, in 2014 and 2019, respectively. He serves in South China Normal University. His research interests include machine learning and computer vision. He has published in TPAMI, CVPR, IJCAI, ACM MM and top AI conference and journals for more than 15 times. 
\end{IEEEbiography}






\end{document}